\documentclass[
 reprint,
 superscriptaddress,
nofootinbib,
 amsmath,amssymb,
 aps,prc
floatfix,
]{revtex4-2}

\usepackage{graphicx}
\graphicspath{{figures/}} 
\usepackage{dcolumn}
\usepackage{bm}

\usepackage{microtype}

\usepackage[hyperindex=true,colorlinks=true]{hyperref}

\hypersetup{citecolor=[rgb]{0.125,0.679,0.196}} 
\usepackage{url}

\usepackage{amsmath}
\usepackage{amssymb}
\usepackage{braket}
\usepackage{mathrsfs}
\usepackage{bbm}
\usepackage{stmaryrd} 
\usepackage{slashed}
\usepackage{bm} 
\usepackage{color} 

\usepackage{physics}
\usepackage{float}
\usepackage{graphicx}

\usepackage{tikz}
\usetikzlibrary{decorations.pathreplacing}
\usepackage[absolute,overlay]{textpos}
\usetikzlibrary{shapes,arrows}

\tikzstyle{decision} = [diamond, draw, fill=blue!20, 
text width=4.5em, text badly centered, node distance=3cm, inner sep=0pt]
\tikzstyle{block} = [rectangle, draw, fill=blue!20, 
text width=10em, text centered, rounded corners, minimum height=4em]
\tikzstyle{line} = [draw, -latex']
\tikzstyle{cloud} = [draw, ellipse,fill=red!20, node distance=3cm,
minimum height=2em]
\usetikzlibrary{shapes.misc,positioning,fit,decorations.pathreplacing, calligraphy}
\usetikzlibrary{positioning}
\usetikzlibrary{arrows,decorations.markings}

\newcommand*{\transpose}{^{\mkern-1.5mu\mathsf{T}}} 
\def\orcid#1{\kern .08em\href{https://orcid.org/#1}{\includegraphics[keepaspectratio,width=0.7em]{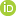}}}

\usepackage{placeins}

\begin{document}

\title{Machine learning one-dimensional spinless trapped fermionic systems with neural-network quantum states}

\author{J. W. T.~Keeble \!\!\orcid{0000-0002-6248-929X}}
\email{j.keeble@surrey.ac.uk}
\affiliation{
 Department of Physics, University of Surrey, Guildford GU2 7XH, United Kingdom
 }

\author{M.~Drissi \!\!\orcid{0000-0001-9472-6280}}
\email{mdrissi@triumf.ca}
\affiliation{
 Department of Physics, University of Surrey, Guildford GU2 7XH, United Kingdom
 }
\affiliation{TRIUMF, 4004 Wesbrook Mall, Vancouver, BC V6T 2A3, Canada}

\author{A.~Rojo-Francàs \!\!\orcid{0000-0002-0567-7139}}
\affiliation{Departament de F\'isica Qu\`antica i Astrof\'isica (FQA),
Universitat de Barcelona (UB), 
c. Mart\'i i Franqu\`es 1, 08028, Barcelona, Spain}

\affiliation{Institut de Ci\`encies del Cosmos (ICCUB), 
Universitat de Barcelona (UB), 
c. Mart\'i i Franqu\`es 1, 08028, Barcelona, Spain}

\author{B.~Juliá-Díaz \!\!\orcid{0000-0001-8658-6927}\, }
\affiliation{Departament de F\'isica Qu\`antica i Astrof\'isica (FQA),
Universitat de Barcelona (UB), 
c. Mart\'i i Franqu\`es 1, 08028, Barcelona, Spain}

\affiliation{Institut de Ci\`encies del Cosmos (ICCUB), 
Universitat de Barcelona (UB), 
c. Mart\'i i Franqu\`es 1, 08028, Barcelona, Spain}

\author{A.~Rios \!\!\orcid{0000-0002-8759-3202}}
\affiliation{
 Department of Physics, University of Surrey, Guildford GU2 7XH, United Kingdom
 }

\affiliation{Departament de F\'isica Qu\`antica i Astrof\'isica (FQA),
Universitat de Barcelona (UB), 
c.~Mart\'i i Franqu\`es 1, 08028, Barcelona, Spain}

\affiliation{Institut de Ci\`encies del Cosmos (ICCUB), 
Universitat de Barcelona (UB), 
c.~Mart\'i i Franqu\`es 1, 08028, Barcelona, Spain}

\date{\today}

\begin{abstract}
We compute the ground-state properties of fully polarized, trapped,
 one-dimensional fermionic systems interacting through a Gaussian potential. 
We use an antisymmetric artificial neural network, or neural quantum state, 
as an
\emph{Ansatz} for the wavefunction and use machine learning techniques to 
variationally minimize the energy of systems from
two to six particles. 
We provide extensive benchmarks for this toy model 
with other many-body methods,
including 
exact diagonalisation and the Hartree-Fock approximation. 
The neural quantum state provides the best energies across a wide range of 
interaction strengths. 
We find very different ground states depending on the sign 
of the interaction. 
In the nonperturbative repulsive regime, the system 
asymptotically reaches crystalline order. In contrast, the strongly 
attractive regime shows signs of bosonization. 
The neural quantum state continuously learns these different phases with 
an almost constant number of parameters and a very modest increase
in computational time with the number of particles.
\end{abstract}

\maketitle

\section{Introduction}

The emergence of machine learning (ML) within science has revolutionized numerous fields, from \emph{ab-initio} quantum chemistry to cosmology, by directly ``learning'' from data to understand physical phenomena~\cite{Carleo2019}. 
Learning algorithms based on neural networks are underpinned by universal approximation theorems (UATs), which allow for, in principle,
an arbitrary accuracy of data representation. 
UATs are, however, \emph{existence} theorems, which state that such benefits are possible, but do not necessarily indicate \emph{how} such benefits are achieved \cite{cybenko1989approximation,hornik1991approximation,irie1988capabilities,kolmogorov1957representation}.
This allows for a wide range of research to build ML-based algorithms that represent physical phenomena,
most notably the quantum many-body problem. 

The first use of ML in a quantum many-body system was pioneered in Ref.~\cite{Carleo2017}   
and focused on discrete systems. Since then,
several different physical systems have been tackled
with these techniques~\cite{Saito2017,Saito2018,choo2018symmetries,Luo2019,Rigo2022,Fore2022,Mujal2021,Saraceni2020}. Applications
in quantum chemistry, like FermiNet~\cite{Pfau2020,Spencer2020}
or PauliNet~\cite{Hermann2020,Schaetzle2021}, 
have paved the way for
accurate solutions of the electronic Schr\"odinger equation~\cite{Wilson2021,Wilson2022,vonGlehn2022,Scherbela2022,Luo2023}. The marriage of quantum states and neural networks has lead to
the novel field of neural-network quantum states (NQSs) \cite{choo2018symmetries}.

ML-based NQS approaches solve the Schr\"odinger equation variationally by representing the wavefunction as a neural network~\cite{Carleo2017}.
NQSs are formulated so that they explicitly respect the symmetry of the many-body wave function, e.g., antisymmetry in the case of fermions.
So far, there are indications that NQSs can compress the relevant information of many-body wave functions in a compact way~\cite{Carleo2017, Saito2018,Pfau2020,Spencer2020,Hermann2020,Hermann2022,Choo2020,netket3_2021}.
While NQSs come in many different setups,  
first-quantized, real-space formulations are often employed 
to perform integrals over many-particle variables,
using Monte Carlo (MC) techniques. 
This approach essentially boils down to variational Monte Carlo (VMC) \cite{ceperley1977monte} with more expressible \emph{Ans\"atze}.
Moreover, the use of a ML framework allows for efficient updates of NQS 
parameters via automatic differentiation (AD) techniques \cite{paszke2017automatic}. Having direct access to the many-body wave function has the added benefit of allowing, in principle, the calculation of any many-body observable and, 
potentially, the simulation of many-body dynamics~\cite{Carleo2017}.
 
Our focus here is on providing a minimal implementation of a NQS for solutions of the many-body Schr\"odinger equation in a simple system
of potential interest in condensed matter. 
As a proof of concept, we turn our attention to one-dimensional systems, 
which are computationally simpler than three-dimensional ones,
and fully polarized (or spinless) fermions. 
We assume the fermions to be in a harmonic trap and their pair-wise interactions are characterized via a finite-range interaction. This exploratory work acts as a stepping stone towards the creation of NQSs that describe more complex nuclear systems~\cite{Keeble2020,Keeble_thesis,Adams2021,Gnech2022,Lovato2022}.

Under certain circumstances, bosons and fermions can hold similar properties.
This phenomenon is referred to as the Fermi-Bose duality and is particularly prominent in one spatial dimension settings~\cite{Girardeau1960relationship,Valiente2020}. The duality manifests itself with strongly interacting fermions acting like weakly interacting bosons and vice versa~\cite{Girardeau2004}. Traditionally, the duality has been
discussed in terms of spin-$1/2$ particles with contact interactions~\cite{Sowinski2019,Minguzzi2022}. 
In the case of fully polarized fermions, the Pauli exclusion principle restricts our wave function \emph{Ansatz} to be antisymmetric with respect to particle position, which results in the physical effect of forbidding S-wave interactions. The corresponding interactions primarily consist of P-wave, odd parity terms~\cite{Girardeau2003,Girardeau2004,Girardeau2004b,Koksik2018,Koksik2020}.
We build a toy model here, which neglects some physically relevant information by utilizing an even-parity interaction, 
 and we discuss benchmarks between several different methods to ascertain the quality of the NQS \emph{Ansatz}. 

Interestingly, the considered quantum many-body 
problem is nowadays within reach experimentally 
in ultracold atomic laboratories 
worldwide~\cite{bloch2008, giorgini2008}. Starting from the 
production of degenerate Fermi gases~\cite{demarco1999}, experimentalists are able to tune the interactions among fermions and study the equation of state~\cite{navon2010} and even produce few-fermion systems in a controlled 
way~\cite{wenz2013}. 

In this paper, we benchmark a minimalist implementation of the FermiNet NQS in Ref.~\cite{Pfau2020} to represent 
one-dimensional trapped fermionic systems. In Sec.~\ref{sec:system},
we define the system and the toy model Hamiltonian we consider. 
We provide a detailed explanation of the NQS and the benchmark many-body methods
in Sec.~\ref{sec:methods}. A detailed analysis of the obtained results is provided in \ref{sec:results}. We provide conclusions and an outlook of future research in 
Sec.~\ref{sec:conclusion}.

\section{System and Hamiltonian}
\label{sec:system}

We study a system of $A$ identical fermions with mass $m$ trapped in a harmonic trap 
of frequency $\omega$. We neglect spin in the following, assuming that the system is fully polarized. 
We focus on a finite-range Gaussian interparticle interaction, which yields the following Hamiltonian in real space:
\begin{align}
\hat H =& - \frac{\hbar^2}{2m} \sum_{i=1}^A  \nabla_i^2 + \frac{1}{2} m \omega^2 \sum_{i=1}^A  x_i^2 
\nonumber \\
&+ 
\frac{V}{\sqrt{ 2 \pi} \sigma} \sum_{i<j}  \exp[ - \frac{ (x_i-x_j )^2}{2 \sigma^2} ] \, .
\label{eq:hamiltonian2}
\end{align}
The Gaussian interaction is characterized by an interaction strength, $V$, and an interaction range, $\sigma$.
We choose these so that, in the limit $\sigma \to 0$ the potential becomes a contact interaction, $\to V \delta( x_i - x_j )$.

We use harmonic oscillator (HO) units throughout the remainder of this work:
lengths are defined in terms of $a_\text{ho} = \sqrt{ \hbar/ m \omega }$ and energies are measured in units of $\hbar \omega$. 
The Hamiltonian becomes
\begin{align}
\hat H =& - \frac{1}{2} \sum_{i=1}^A  \nabla_i^2 + \frac{1}{2} \sum_{i=1}^A  x_i^2  + 
\frac{V_0}{\sqrt{ 2 \pi} \sigma_0} \sum_{i<j}  \exp[ - \frac{ (x_i-x_j )^2}{2 \sigma_0^2} ] .
\label{eq:hamiltonian}
\end{align}
The interaction range is redefined, so that
$
\sigma_0 =  \sigma / a_\text{ho} \, .
$
The dimensionless interaction strength $V_0$ is related to the dimensionful constant $V$ by
$
V_0 = V / (a_\text{ho} \hbar \omega ).
$

For spinless fermions, the presence of a finite range interaction is a necessary
condition to observe interaction effects. Indeed, without spin, the many-body wave function is 
antisymmetric on the space variables $\{ x_i , i=1,\dots,A \}$. For pairs of particles $i \neq j$,
the wave function cancels whenever $x_i=x_j$. As a consequence, contact interactions do not
contribute to the energy of the system.  
Na\"ively, one expects such interaction
effects to be relatively small compared with interactions of the same strength in the spinful case. 

The specific choice of a Gaussian form factor for the interaction 
is dictated mostly by practical reasons~\cite{bloch2008,Sowinski2019}.
First and foremost, because of the simplicity of the associated integrals,
Gaussians can be easily handled in many-body simulations, including exact diagonalization~\cite{RojoFrancas2020} as well as stochastic methods~\cite{Varga}.
Second, as already noted above, normalized Gaussian interactions can be used to
approach the contact-interaction limit by tuning the range parameter $\sigma \to 0$. 
Third, our ultimate goal is to simulate nuclear physics systems. 
Finite-range interactions are particularly relevant for
nuclear physics applications, where the range of the interaction is related to 
the mass of meson force carriers~\cite{RingSchuck1980}. 
There are examples 
of nuclear interactions with a Gaussian (or a sum of Gaussians) 
form factor, like
the Gogny force~\cite{RingSchuck1980}. 
We also note that Gaussian interactions have been extensively 
used in the analysis of several many-body systems, including the 
so-called ``Gaussian characterization" of 
universal behavior~\cite{Kievsky21}. 

Previous literature on spinless fermions has employed odd-parity interactions, which in the zero-range
limit behave as $V(x) \approx \overleftarrow{\partial_x} \delta(x) \overrightarrow{\partial_x}$. This is the correct limit for the 
interaction of spinless fermions and it cannot be reproduced with
a Gaussian potential. As such, our results should be considered as 
an academic toy model, and for this reason we focus on methodological 
benchmarks as opposed to experimental applications. 
We leave the study of odd parity potentials within an NQS framework for further study.

\subsection{Noninteracting case}

Before providing more details on how interactions
are considered, we turn our attention briefly to the analytically solvable, noninteracting
case. In the absence of spin, each fermion can occupy 
a single-particle level with energy $\epsilon_n=n+1/2$ characterized by a single
quantum number $n$. 
The single-particle wavefunctions are HO eigenstates, 
\begin{align}
\varphi_n(x) &= \mathcal{N}_n e^{- \frac{x^2}{2} } H_n \left( x \right) \, ,
\label{eq:HOwfs}
\end{align}
with $H_n(x)$ the $n$th Hermite polynomial and 
$\mathcal{N}_n = 1/\sqrt{2^n n! \sqrt{\pi} }$, 
a normalization constant.
The many-body wave function is a pure Slater determinant. It factorizes into an overall Gaussian envelope times a determinant involving only Hermite polynomials:
\begin{align}
\Psi(x_1, \ldots, x_A) =& \frac{1}{\sqrt{A!}} 
\left[ \Pi_{n=0}^{A-1} \mathcal{N}_n e^{- \frac{x_i^2}{2} }  \right] \times
\label{eq:determinant}
\\
&\begin{vmatrix} 
H_0\left( x_1 \right) & H_0\left( x_2 \right)  & \ldots & H_0\left( x_A \right)  \\
H_1\left( x_1 \right) & H_1\left( x_2 \right)  & \ldots & H_1\left( x_A \right)  \\
\vdots & \vdots & \ddots & \vdots \\
H_{A-1}\left( x_1 \right) & H_{A-1}\left( x_2 \right)  & \ldots & H_{A-1}\left( x_A \right)  
\end{vmatrix} \nonumber.
\end{align}
This, in turn, may be further simplified in terms of 
Vandermonde determinants~\cite{Sowinski2019,Lubos2006}.

The total energy of the many-body system is easily obtained by adding up all the occupied single-particle state energies,
\begin{align}
E_A = \sum_{n=0}^{A-1} \epsilon_n =\frac{A^2}{2}  \, ,
\label{eq:E_A_noninteracting}
\end{align}
and it scales with $A^2$.
These noninteracting benchmarks are useful, especially in setting up the NQS \emph{Ansatz}. In particular, as we explain below, we pretrain neural networks to the noninteracting solution. This provides an initial confined, stable and physical result from which we can start the relatively demanding variational simulations.

\subsection{Density matrices}

We can further characterize correlations in the system by employing
many-body density matrices~\cite{Lowdin1955,Knight2022}.  
The one-body density matrix (OBDM) for the $A$-body system is 
defined as the following
$A-1$ integral over the many-body wave function,
\begin{align}
    \rho(x_1',x_1) = A \int dx_2 \ldots dx_A 
    &\Psi^*(x_1', x_2, \ldots, x_A) \nonumber \\  
    \times &\Psi(x_1, x_2, \ldots, x_A) .
    \label{eq:OBDM}
\end{align} 
The diagonalization of $\rho$ in the space representation,
\begin{align}
    \int d \bar x \rho(x,\bar x) \phi_\alpha (\bar x) &= n_\alpha \phi_\alpha(x), 
    \label{eq:occ_numbers}
\end{align} 
gives rise to the so-called natural orbitals, $ \phi_\alpha(x)$, 
as well as the
occupation numbers $n_\alpha$.
$\alpha$ is a discrete index running, in principle, from $\alpha=0$ to 
infinity.
The spectral decomposition of the OBDM allows for the following expansion:
\begin{align}
\rho(x_1',x_1) = \sum_{\alpha=0}^\infty n_\alpha 
\phi^*_\alpha(x_1') \phi_\alpha(x_1).
\label{eq:OBDM_general}
\end{align}
We work with a normalization such that 
$\sum_\alpha n_\alpha =A$.

In a noninteracting or a Hartree-Fock (HF) ground-state, 
one finds
\begin{align}
n_\alpha &= 
\begin{cases}
1, \quad \alpha < A ,\\
0, \quad \alpha \ge A.
\end{cases}
\label{eq:occupations_free}
\end{align}
The sum in Eq.~(\ref{eq:OBDM_general}) is thus naturally truncated to $A$ terms.
In addition, for the noninteracting case, 
the natural orbitals correspond to the single-particle
states of Eq.~(\ref{eq:HOwfs}). 
Performing the sum for a noninteracting system, one finds that 
the OBDMs have the form
\begin{align}
\rho(x_1',x_1)&= \frac{e^{-\frac{x_1'^2}{2}-\frac{x_1^2}{2}}}{\sqrt{\pi}}
\mathcal{R}^{(A)}(x_1',x_1) ,
\label{eq:OBDM_HO}
\end{align}
where $\mathcal{R}^{(A)}(x_1',x_1)$ 
is a polynomial of at most order $A-1$ in both $x_1$ and $x_1'$~\cite{Schilling2013,Schilling2016,Rius2023}. 

One can also prove that a system with the occupation numbers 
described by Eq.~(\ref{eq:occupations_free}) has a Slater determinant as 
a many-body wave function~\cite{Gross1991}.
In other words, deviations from the uncorrelated values $n_m=1$ and $n_m=0$ 
provide a solid metric for intrinsic correlations in the system~\cite{Koksik2020}. 

The two-body density matrix is also an excellent indicator of 
intrinsic correlations.
It is usually defined as~\cite{Lowdin1955},
\begin{align*}
    \Gamma(x_1',x_2'; x_1,x_2) = 
    {A \choose 2}
    \int dx_3 \ldots  dx_A 
    &\Psi^*(x_1', x_2', \ldots, x_A) \nonumber \\  
    \times &\Psi(x_1, x_2, \ldots, x_A) \, .
\end{align*} 
The positive-definite diagonal elements of this matrix provide the pair correlation
function (PCF) of the system, 
\begin{align}
    g(x_1,x_2) = \Gamma(x_1'=x_1,x_2'=x_2; x_1,x_2) \, ,
    \label{eq:g_pair_dist}
\end{align} 
which has a direct physical interpretation in terms of the probability
of finding a particle at position $x_1$ when another one lies at $x_2$~\cite{Knight2022}. 
Closed expressions can be found for this object in the noninteracting case, too. 
In this case, the correlation function has a structure
of the type
\begin{align}
g(x_1,x_2)&= \frac{e^{-x_1^2 -x_2^2 }}{\pi} (x_1-x_2)^2
\mathcal{G}^{(A)}(x_1,x_2) ,
\label{eq:correlation_function_HO}
\end{align}
where $\mathcal{G}^{(A)}(x_1,x_2)$ 
is again a polynomial of at most order $2(A-2)$ in both 
$x_1$ and $x_2$~\cite{Rius2023}. 

\section{Methods}
\label{sec:methods}

We now describe the different quantum many-body approaches that we have used to benchmark 
our few-body solutions. We start by providing a description of the NQS \emph{Ansatz}, and then move on to 
describe briefly the exact diagonalization method and the 
Hartree-Fock approximation for 
the cases with $A>1$. 
We end the section with a description of two additional numerical 
methods employed to benchmark specifically the $A=2$ case. 

\subsection{Neural quantum states}
\label{subsec:NQS}

\subsubsection{Architecture}

\begin{figure*}
\includegraphics[width=\linewidth]{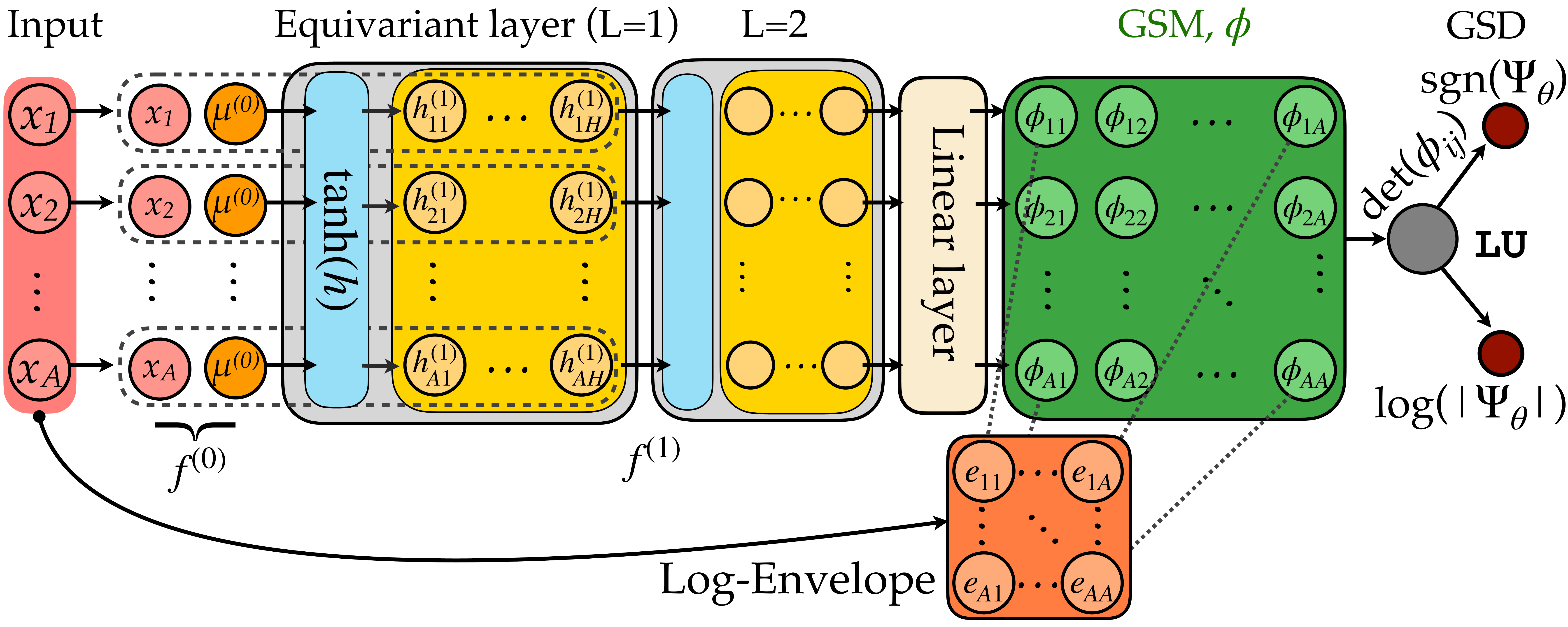}
	\caption{The NQS \emph{Ansatz} of this work for $L=2$ equivariant layers of $H$ hidden nodes. 
	The input to the network are the $\{x_i, i=1, \ldots, A\}$ particle positions.
	These are processed by $2$ equivariant layers (light gray areas),
	to ensure that the NQS maintains equivariance throughout its forward pass. 
	The grey dashed lines denote the row-wise application of the shared layer.
    Its nonlinear activation function is an hyperbolic tangent (blue area).
    The output of the second intermediate equivariant layer $h^{(2)}$ is passed through
    a shared linear layer to yield a
	$A \times A$ matrix $M$ which is element-wise multiplied with the log-envelope layer (orange area)
    thus yielding our GSM $\phi$ (green area).
    In a final step, a LU decomposition is used to find the 
	sign and logarithm of the absolute value of the many-body wavefunction, $\Psi_\theta$.}
	\label{fig:deep-wavefunction-ansatz}
\end{figure*}

Our NQS \emph{Ansatz} is a fully antisymmetric neural network, inspired by the  
implementation of FermiNet~\cite{Pfau2020}. 
The input to our network are the $A$ positions of fermions in the system, $\{ x_i , i = 1, \ldots, A \}$,
and the output is the
many-body wave function $\Psi_\theta(x_1,\ldots,x_A)$,
which depends on a series of network weights and biases,
succinctly summarized by a multidimensional variable $\theta$.
We provide a schematic representation of the network architecture in Fig.~\ref{fig:deep-wavefunction-ansatz}.
The network is composed of 4 core components:
equivariant layers, generalized Slater matrices (GSMs), log-envelope functions, 
and a summed signed-log determinant function.
We now proceed to describe each of the 4 core components of the NQS.

In order for the NQS to respect antisymmetry,
we enforce permutation equivariance across the network. 
The inputs to our network are an $A$-dimensional vector, $x \in \mathbb{R}^{A}$. The outputs of
an equivariant layer, $h_i$, are such that any permutation of the input variables, 
$\pi$, permutes the outputs too,
$h_i( x_{\pi(1)}, \ldots, x_{\pi(A)} ) = h_{\pi(i)} ( x_1, \ldots, x_A )$.

To ensure equivariance, we follow the methodology of Ref.~\cite{sannai2019universal}. 
We preprocess the input layer by adding a permutation-invariant feature,
which in our case is the mean many-body position,
$\mu^{(0)} = \frac{1}{A} \sum_{i=1}^A x_i$. 
The input to the first layer is then a concatenation of $x \in \mathbb{R}^{A}$, and 
the corresponding mean position, $\mu^{(0)} \in \mathbb{R}$. 
This defines an input feature $f^{(0)} \in \mathbb{R}^{A \times 2}$, such that
\begin{equation}
    f^{(0)}_{ij} =
        \begin{cases}
            x_i & \text{if } j = 1 \\
            \mu^{(0)} & \text{if } j = 2 \ .
        \end{cases}
\end{equation}

The first equivariant layer $(L=1)$ of the network is shown
in a grey area in Fig.~\ref{fig:deep-wavefunction-ansatz}. 
Each row of the input feature is passed through a 
shared layer to build an intermediate $H$-dimensional
representation of the positions, $h^{(1)} \in \mathbb{R}^{A \times H}$
(yellow area in Fig.~\ref{fig:deep-wavefunction-ansatz}).
This layer consists of a linear transformation with weights $W^{(1)} \in \mathbb{R}^{H \times 2}$
and biases $b^{(1)} \in \mathbb{R}^{1 \times H}$ combined with a nonlinear activation function.
We choose a hyperbolic tangent activation function, since this is continuous and
differentiable, a requirement when it comes to 
computing many-body kinetic energies.
Explicitly, each row $h^{(1)}_{i} \in \mathbb{R}^{1 \times H}$ reads 
\begin{equation}
    h^{(1)}_{i} = \tanh\left(f^{(0)}_i  W^{(1)\transpose} + b^{(1)}\right) \ .
\end{equation}
The second equivariant layer $(L=2)$ takes the 
output of the first layer, $h^{(1)}$, and their column-wise averages $\mu^{(1)} \in \mathbb{R}^{H}$,
to define a new input feature $f^{(1)} \in \mathbb{R}^{A \times 2H}$ such that
\begin{equation}
    f^{(1)}_{ij} =
        \begin{cases}
            h^{(1)}_{ij} & \text{if } j \leq H \\
            \mu^{(1)}_{i} & \text{if } j > H \ .
        \end{cases}
\end{equation}
Similarly to the first layer, the input feature is passed through a shared layer with
weights $W^{(2)} \in \mathbb{R}^{H \times 2H}$, biases $b^{(2)} \in \mathbb{R}^{1 \times H}$
and the same nonlinear activation function.
A residual connection is also added so that each row of the output
$h^{(2)}_i \in \mathbb{R}^{1 \times H}$ explicitly reads
\begin{equation}
    h^{(2)}_{i} = \tanh\left(f^{(1)}_i  W^{(2)\transpose} + b^{(2)}\right) + h^{(1)}_i \ .
\end{equation}
After the second equivariant layer, $h^{(2)}$ goes through a linear layer
(see beige area in Fig.~\ref{fig:deep-wavefunction-ansatz})
with shared weights $W^{(M)} \in \mathbb{R}^{A \times H}$
and biases $b^{(M)} \in \mathbb{R}^{1 \times A}$ to output a matrix
$M \in \mathbb{R}^{A \times A}$.
Each row $M_i \in \mathbb{R}^{1 \times A}$ reads
\begin{equation}
    M_i = h^{(2)}_i W^{(M)\transpose} + b^{(M)} \ .
\end{equation}

At this stage, asymptotic boundaries conditions,
such as the wave function decaying at infinity,
are not yet incorporated. 
To this end, we employ an envelope function \cite{martin_reining_ceperley_2016}, implemented in the log-domain for numerical stability. The log-envelope matrix has the form
\begin{align}
    \ln( e_{ij} )= -\left(x_i W^{(e)}_j \right)^{2},
    \label{eq:envelope}
\end{align}
where $i$ and $j$ represent the index for the particle and the orbital, 
respectively, and $W^{(e)} \in \mathbb{R}^{A}$ is the weight that is learned 
to determine the log-envelope of each orbital. 
We use Gaussian envelopes, instead of the exponential ones of Ref.~\cite{Pfau2020}, which are closer to the noninteracting
solution of Eq.~(\ref{eq:determinant}).

We then take an element-wise product of $M$ with
the corresponding envelope,
\begin{align}
    \phi_{ij} = M_{ij} e_{ij}, 
\end{align}
which yields a GSM. 
Each element of this matrix, 
$\phi_{ij} = \phi_j (x_i ; \{x_{\slash i} \})$,  
may be understood as a 
generalized single-particle orbital on state $j$. This
orbital does not only depend on 
the position of the particle $i$, but also on the positions of other particles 
in a permutation-invariant way, as indicated by the notation 
$\{x_{\slash i} \}$~\cite{Pfau2020}.
This has the significant benefit of making all orbitals depend on \emph{all} particle positions,
which amounts to an efficient encoding of backflow correlations in the
wavefunction~\cite{Luo2019,Pfau2020,becca_sorella_2017}, 
as discussed further in Sec.~\ref{subseq:backflow}.
 
Finally, we take the determinant of the GSM $\phi$ to obtain an 
antisymmetric wave function. This is generically referred to as
a generalized Slater determinant (GSD).
In principle, a single GSD is sufficient to represent any antisymmetric
wave function~\cite{hutter2020representing}.
Empirically, we also observe that one GSD
captures nearly all the correlations in this system~\cite{Keeble_thesis}.
We note that the GSD is computed within the log-domain via a 
lower-upper (LU) decomposition. This choice is dictated by numerical stability.  

Our computational framework is general, and we can 
modify both the number of equivariant layers, $L$, and the 
number of hidden nodes, $H$. 
Our NQS uses $L=2$ equivariant layers of $H=64$ hidden nodes each.
This choice shows optimal results in terms of convergence according to 
numerical tests that can be found in Ref.~\cite{Keeble_thesis}. 
Our code can also work with more than one GSM and GSDs. 
If we were to employ $D$ GSDs to describe the system, 
each determinant would have its own envelope 
through a $D$ dependence of the envelope weights $W^{(e)}$ in Eq.~(\ref{eq:envelope})~\cite{Keeble_thesis}. 
In the $D>1$ case, the GSDs are summed via a signed-log-sum-exp function.
We direct the reader to Appendix~A of Ref.~\cite{Keeble_thesis}
for a complete and detailed explanation of the numerical implementation in the $D>1$ case.

\subsubsection{Variational Monte Carlo} 

Having defined the NQS \emph{Ansatz}, we now turn to discussing the details of how we
implement a quantum many-body solution to our problem. 
We solve the Schr\"odinger equation via a VMC approach
in two phases: a \emph{pretraining} to an initial target wavefunction, and an \emph{energy minimization} to the unknown ground-state wave function. 
The pre-training step can be thought of as a supervised learning exercise, where we demand 
that the network reproduces the many-body wave function of the noninteracting system. 
The idea is to obtain an initial state that is physical and
somewhat similar, in terms of spatial extent, to the 
result after interactions are switched on. 
To do so, we minimize the loss,
\begin{align*}
    \mathcal{L}^{\text{Pre}}(\theta)  = \int d x_1 \ldots dx_A &
    \left| \Psi_\theta (x_1, \ldots, x_A) \right|^2 \times \\
    &  \sum_{ij} \left[ 
    \phi_{i} \left(x_j ; \{ x_{\slash j} \} \right) 
    - \varphi_{i}(x_j)  \right]^{2}, \nonumber 
\end{align*}
where $\phi_{i} \left(x_j ; \{ x_{\slash j} \} \right)$
is an element of the GSM 
and $\varphi_i$ is the $i$th HO single-particle state, see Eq.~(\ref{eq:HOwfs}).  
The loss can be reformulated via MC sampling as
\begin{align}
    \mathcal{L}^{\text{Pre}}(\theta) = 
    \mathbb{E}_{X \sim \abs{\Psi_\theta}^{2}} \left[ 
    \sum_{ij} \left[ \phi_{i} \left(x_j ; \{ x_{\slash j} \} \right) - \varphi_{i}(x_j)  \right]^{2}
    \right],
    \label{eq:pre-training-mc}
\end{align} 
where we define $X=(x_1, \ldots, x_A)$ as an $A$-dimensional 
random variable (or walker) distributed according to the Born 
probability of
the many-body wavefunction, $\abs{\Psi_\theta}^{2}$.

Initially, a set of $N_W=4096$
walkers are
independently distributed at the origin of configuration space
using a zero-mean and unit variance $A$-dimensional 
Gaussian distribution. 
The pretraining phase is then iterated for
$10^4$ epochs. 
Each epoch has three different phases.
First, 
the $N_W$ walkers are distributed in proportion to the Born probability of the NQS via a Metropolis-Hastings (MH) algorithm
\cite{metropolis1949monte,hastings1970monte}. 
We run 
$10$ iterations of the MH algorithm per epoch.
Second, we compute the local loss values and back-propagate to
evaluate the gradients of the loss in Eq.~(\ref{eq:pre-training-mc}) 
with respect to the parameters $\theta$.
Third, we update the parameters using the Adam optimizer 
of Ref.~\cite{kingma2014adam}, with default hyperparameters except with a learning rate of 10$\textsuperscript{-4}$. 
Once the pre-training  is complete, the NQS represents an approximate solution to the noninteracting case. 

The next stage of our VMC approach is to minimize the expectation value of
the energy to find the ground-state wave function in a reinforcement learning setting.
Using standard quantum Monte Carlo notation, the expectation value of the energy is formulated as a statistical average over \emph{local} energies, 
\begin{align}
    E(\theta) & = \frac{\langle \Psi_\theta \vert \hat{H} \vert \Psi_\theta \rangle}{\langle \Psi_\theta \vert \Psi_\theta \rangle} =
    \mathbb{E}_{X \sim \abs{\Psi_\theta}^{2}}
    \left[ \Psi_\theta(X)^{-1} \hat{H} \Psi_\theta(X) \right]  .
    \label{eq:stochastic_energy}
\end{align}
Within statistical uncertainties, this expectation value
abides by the variational principle and $E(\theta)$ is larger than
the ground state energy $E_{\textrm{g.s.}}$. 

We compute the kinetic energy in the log domain, which leads to a
local energy 
\begin{align}
E_L(X) &\equiv 
    \Psi_\theta(X)^{-1} \hat{H} \Psi_\theta(X) \nonumber \\
    &=  -\frac{1}{2} \sum_{i=1}^A \left[ 
    \left. \frac{\partial^{2} \ln\abs{\Psi_\theta}}{\partial x_i^{2}} \right|_X + 
    \left. \left(\frac{\partial \ln\abs{\Psi_\theta}}{\partial x_i}\right)^{2} \right|_X \right] \nonumber \\
    &\phantom{=} + \sum_{i=1}^A\frac{x_i^2}{2}  + 
\frac{V_0}{\sqrt{ 2 \pi} \sigma_0} \sum_{i<j}  e^{ - \frac{ (x_i-x_j )^2}{2 \sigma_0^2} } .
\label{eq:local_energy}
\end{align}
The walkers are propagated using a MH algorithm. 
With this stochastic process,  walkers can sample the
arbitrary probability distribution dictated by the wave function
Born probability.
Furthermore, we follow the methodology of Ref.~\cite{Wilson2021} to adapt
on the fly the width of the proposal distribution in the MH sampler.
The aim is to ensure that, on average, approximately 
$50 \%$ of the walkers accept the proposed configuration at each step of the MH algorithm. This allows for the proposal distribution to effectively scale with the system size, leading to a more efficient thermalization of the Markov chain. 

For the sake of numerical stability, we follow Ref.~\cite{Pfau2020} and use an $\ell_1$ norm clipping, in which we calculate a window
of ``acceptable'' local-energy values. 
We choose an $\ell_1$ norm as it is more robust to outliers.
For a batch of local energies, we compute the local energy median, 
$\langle E_L \rangle$, with an associated $\ell_1$ norm deviation 
$\sigma_{\ell_1}$.
The acceptable window is defined as the range
$\langle E_L \rangle \pm 5\sigma_{\ell_1}$.
Any values outside the window are replaced by
the maximally accepted value in each side. 

Once the expectation value of the energy is computed, we update the parameters
of our \emph{Ansatz} using its gradients with respect to $\theta$,
\begin{equation}
    \nabla_{\theta}E(\theta)
        =
        2\mathbb{E}_{X \sim \abs{\Psi_\theta}^{2}}\left[
            \left(E_L(X) - E(\theta)\right)
            \nabla_{\theta}\ln\abs{\Psi_\theta(X)}
        \right] \ .
\end{equation}
The gradient of the energy is computed by applying reverse-mode AD on the auxiliary loss function,
\begin{align}
    \mathcal{L}^{\text{aux}}(\theta)
        &=
        2\mathbb{E}_{X \sim \bot(\abs{\Psi_\theta}^{2})}\left[
            \bot\left( E_L(X) - \mathbb{E}_{X \sim \abs{\Psi_\theta}^{2}}\left[E_L(X)\right]\right)
            \right. \nonumber \\
        &\phantom{= 2\mathbb{E}_{X \sim \bot(\abs{\Psi_\theta}^{2})}[ }
            \left.\vphantom{\left[\bot\left( E_L(X) - \mathbb{E}_{X \sim \abs{\Psi_\theta}^{2}}\left[E_L(X)\right]\right)\right]}
            \times \ln\abs{\Psi_\theta(X)}
        \right] \, ,
    \label{eq:loss-energy}
\end{align}
where $\bot(\ldots)$ denotes the detach function~\cite{zhang2019automatic}. 
The detach function is a common ML implementation tool that allows to combine conveniently
analytical calculations of gradients with an AD algorithm.
When applying AD, the detach function is, by definition, equivalent
to the identity function except that its derivative is equal to zero, i.e.
$\bot(x) = x$ but $\partial \bot(x)/\partial x = 0$.
While the expectation value of Eq.~(\ref{eq:loss-energy}) does not
equal the energy in Eq.~(\ref{eq:stochastic_energy}), 
the derivatives of both expressions with respect to the parameters of the NQS are
equal. The motivation for this auxiliary loss function over the na\"ive implementation
of Eq.~(\ref{eq:loss-energy}) is numerical stability, as it allows for a calculation
avoiding higher-order derivatives.
One can prove that, by exploiting the detach function, 
the highest-ordered derivative that is required to compute the gradient is the 
second-order derivative with respect to $x$ for the kinetic energy. A na\"ive 
application of AD on Eq.~(\ref{eq:stochastic_energy}) to compute the gradient of the energy
would have led to a third-order mixed derivative instead. 

\begin{figure}[t]
\includegraphics[width=\linewidth]{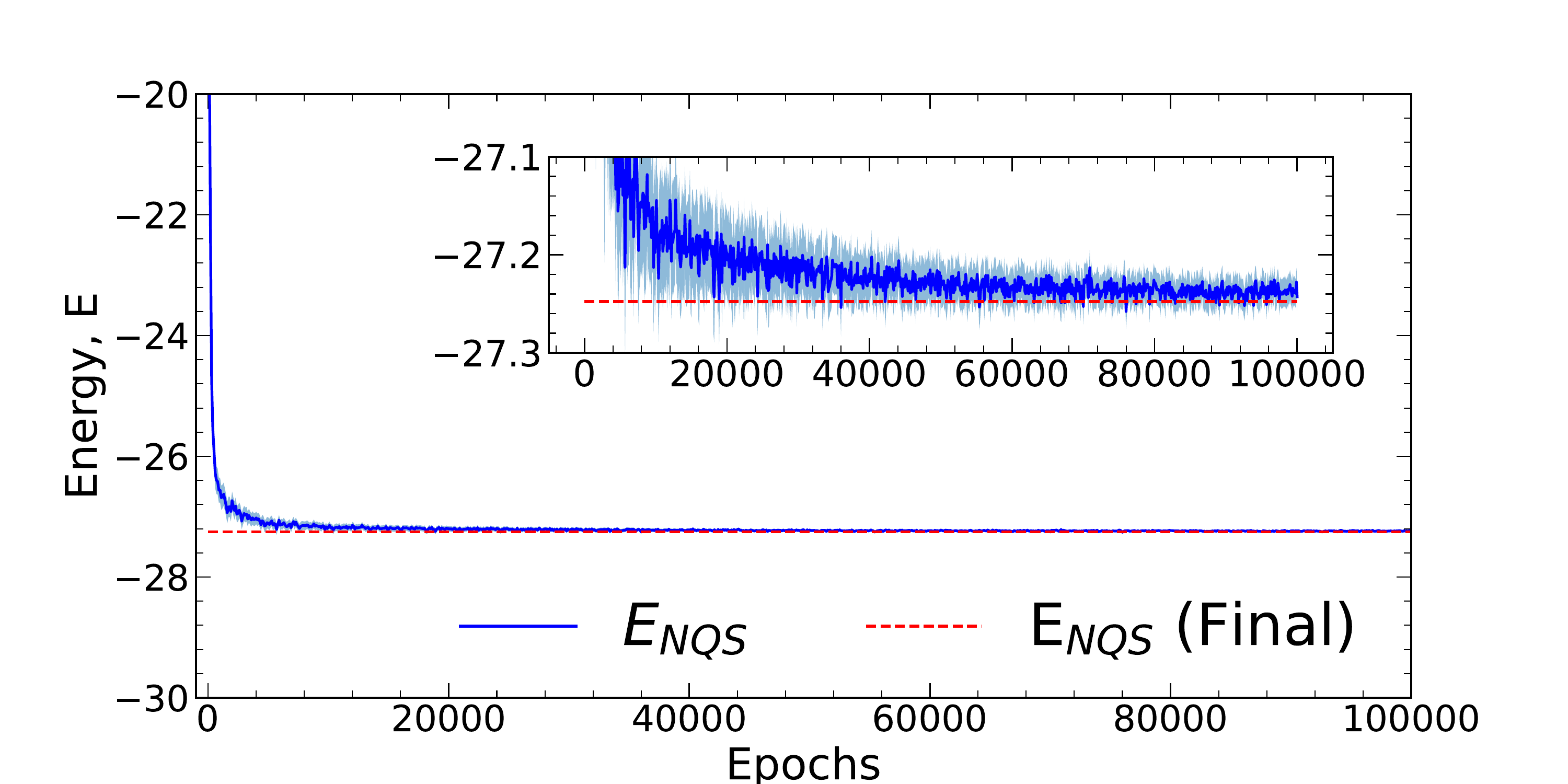}
\caption{
Convergence of the expectation value of the energy for the $A = 6$ system with $V_0=-20$ and $\sigma_0=0.5$, as a function of the epoch. The NQS \emph{Ansatz} is shown as a blue solid line. The 99.7\% (3 standard deviations) confidence interval of the MC energy integral is shown with a shaded blue region.
The final energy of the NQS, calculated via the blocking method, is shown in the red dashed line. For clarity, we show every 100\textsuperscript{th} epoch in the convergence.
}
\label{fig:convergence}
\end{figure}

Just as in the pretraining phase, we use $N_W=4096$ walkers and minimize the energy for 
$10^5$ epochs. Each epoch starts with $10$ MH sampling steps, computes the local energies
through Eq.~(\ref{eq:local_energy}), and back-propagates the local gradients of Eq.~\eqref{eq:loss-energy}
in order to converge the NQS towards the ground-state wave function.
The parameters are updated using the Adam optimizer with default hyperparameters except with a learning rate of 10$\textsuperscript{-4}$~\cite{kingma2014adam}.
We show in Fig.~\ref{fig:convergence} the 
evolution of the energy for one of the
minimization runs, for $A=6$, $V_0=-20$
and $\sigma_0=0.5$. This represents one of the
most challenging cases presented here, with a 
relatively large number of particles and a 
strong deviation from the noninteracting case.
We show both the average value of $E$ (solid
line) and the 99.7\% confidence interval of the 
integral of Eq.~(\ref{eq:stochastic_energy}). 
We find that in about $10 000$ epochs the
energy is within $0.15 \%$ of the final value. 
The energy converges towards the ground state
steadily, with relatively small oscillations
around a central value (see inset of Fig.~\ref{fig:convergence}).

We assume that the NQS reaches the ground state after $10^5$ epochs and freeze the 
parameters $\theta$ at that stage. 
To compute a final estimate of the energy, we follow Ref.~\cite{scemama2006efficient} 
and calculate it over multiple batches via the ``blocking'' method. 
We use $10^4$ batches of $4096$ walkers each, for a total of $\approx 4.1 \times 10\textsuperscript{7}$ samples. 
We remove the aforementioned $\ell_1$ norm clipping, so as to not bias the final energy measurement. The final energy (and its standard deviation) is taken by averaging over these measurements.
The total standard deviation is also obtained employing the blocking method of Ref.~\cite{scemama2006efficient}. 
The standard deviation of an individual chain is defined as,
\begin{align}
    \sigma_{B}^2 & = \sigma^2 / (L_B / N_{corr} ),
\end{align}
with $\sigma_{B}$ and $\sigma$ being the empirical standard deviation of a block $B$ and of all samples, respectively. $L_B$ is the length of a given block $B$ with correlation length
$N_{corr}$.
This standard deviation is usually extremely small and does not affect our conclusions.
All the observables throughout this paper are computed with this superset of samples, which allows for sufficiently accurate measurements.
More details on the implementation of our NQS can be found in Ref.~\cite{Keeble_thesis} as well as the GitHub repository \cite{jwtkeeble2023}.

\subsubsection{Scaling with particle number}

We now briefly comment on the scaling of the NQS method with 
the number of particle $A$.
To asses the scaling, we estimate the memory and time requirements of
the four main layers of our NQS in their forward pass.
The first two layers (gray areas in Fig.~\ref{fig:deep-wavefunction-ansatz}) 
have, including biases, $H\times(2+1)$ and $H\times(2H+1)$ parameters, respectively. 
We stress that the number of parameters here is independent of the particle number.
These two layers are shared and applied on $A$ row-vectors, each corresponding to a different particle, 
so the
size of those layers is constant in $A$, while the time complexity is linear in $A$.
The two remaining layers, the GSM and log-envelope layers, have,
including biases, $D\times (H+1) \times A$
and $D \times A$ parameters, respectively.
For these two layers, the size is linear in $A$, while the time complexity
is quadratic in $A$.
Lastly, the final evaluation of the determinants of the GSMs is parameter-free, but with a time complexity
scaling as $D \times A^3$.

In practice, we have set $H=64$, $L=2$ and $D=1$ while $A$ varies from two to six.
Memory-wise, the dominant cost is the storage of the intermediate
layer weight matrix (quadratic in $H$) despite being constant in $A$.
For example, the $A=2$ NQS has $8580$ parameters
and the $A=6$ NQS has $8844$ parameters, which corresponds to a small 
$3\%$ increase due to the linear dependence on $A$ of the number of parameters. 
Regarding the computational cost, the asymptotically dominant term
is the evaluation of the determinants (cubic in $A$ for the forward pass).
In our simulations, however, we find that the increase of walltime per epoch with
$A$ for our NQS is relatively small for the system sizes studied.
In the $A = 2$ case, the walltime is approximately $0.1$~s per 
epoch, whereas it raises to $0.3$~s per epoch in the $A = 6$.
This linear scaling suggests that we have not yet reached the asymptotically large $A$ behavior. 
Further profiling of our VMC simulations indicates that, for $A \leq 6$,
the dominant contribution in terms of walltime actually comes from the MH 
sampling~\cite{Azzam2023}. This part of the VMC algorithm is significantly more time consuming
than any of the forward or backward network passes, and hence should be addressed in any
time optimisation strategy. 
For reference, we have run all our experiments on a single P4000 Quadro GPU
and overall walltime may be significantly sped up with hardware improvements. 
 
Finally, in all the above considerations we have assumed $H$ is kept fixed while $A$ increases.
The mild scaling laws that are resulting from this assumption are of course limited
by the degradation of the expressivity of the network when $A$ increases.
As a minimal requirement one must have $A \leq H$
to ensure the that the GSMs are full-rank and the NQS is not vanishing.
To test how the expressivity of the network evolves with $A$, we have performed
some additional numerical tests for the $A=12$ system with the same
values of $L$ and $D$, but changing $H$. While we do not have exact benchmarks to
compare with, we have found that, in this case, the NQS \emph{Ansatz} still reaches an energy
lower than the corresponding HF upper bound for $H > 16$. 
The final energy values for $H=64$, $128$ and $256$ are essentially converged.
We interpret this as a sign that the 
network expressivity can still be kept under control even if an increase of $H$
might be needed to ensure that the NQS bias remains negligible. Further explorations for larger particle numbers
need to be carried out to better characterize any expressivity 
limits in our NQS \emph{Ansatz}.

\subsubsection{Backflow correlations}
\label{subseq:backflow}

Our approach is different to previous  ML implementations~\cite{Pfau2020,Adams2021,Gnech2022,Lovato2022},
particularly FermiNet, in various ways.
We drop the convolution layers,
omit any accommodation of spin dependency,
use Gaussian envelopes instead of exponential ones,
and work with improved numerical stability on the sum of GSDs. 
Crucially, we also do not incorporate any explicit Jastrow factor.
This does not hamper the quality of the corresponding wave function, as we shall see below. 

The most relevant aspect underlying the success of our NQS implementation is its ability to incorporate backflow correlations. 
In the noninteracting and the HF descriptions, 
the many-body wave function is a Slater determinant of orbitals that  depend on single-particle positions alone~\cite{martin_reining_ceperley_2016,Blaizot1986,Gross1991}. 
In our NQS implementation, we exploit the permutation equivariance together with a determinant layer, which introduces a form of backflow to the wavefunction representation~\cite{Feynman_backflow,Holzmann2006,Holzmann2019}. 
This converts single-particle orbitals into quasi-particle orbitals with an extended, permutation invariant dependence on the position of all particles, and allows for a shift of the nodal surface of the NQS wave function~\cite{Ceperley1991}. 
Backflow transformations allow to incorporate correlations into the many-body wave functions. Our NQS representation, in particular,  provides a flexible, numerically cheap and relatively universal representation of these correlations~\cite{Luo2019,Pfau2020}, 
which can be efficiently learned during the VMC process.

\begin{figure}[t]
    \includegraphics[width=\linewidth]{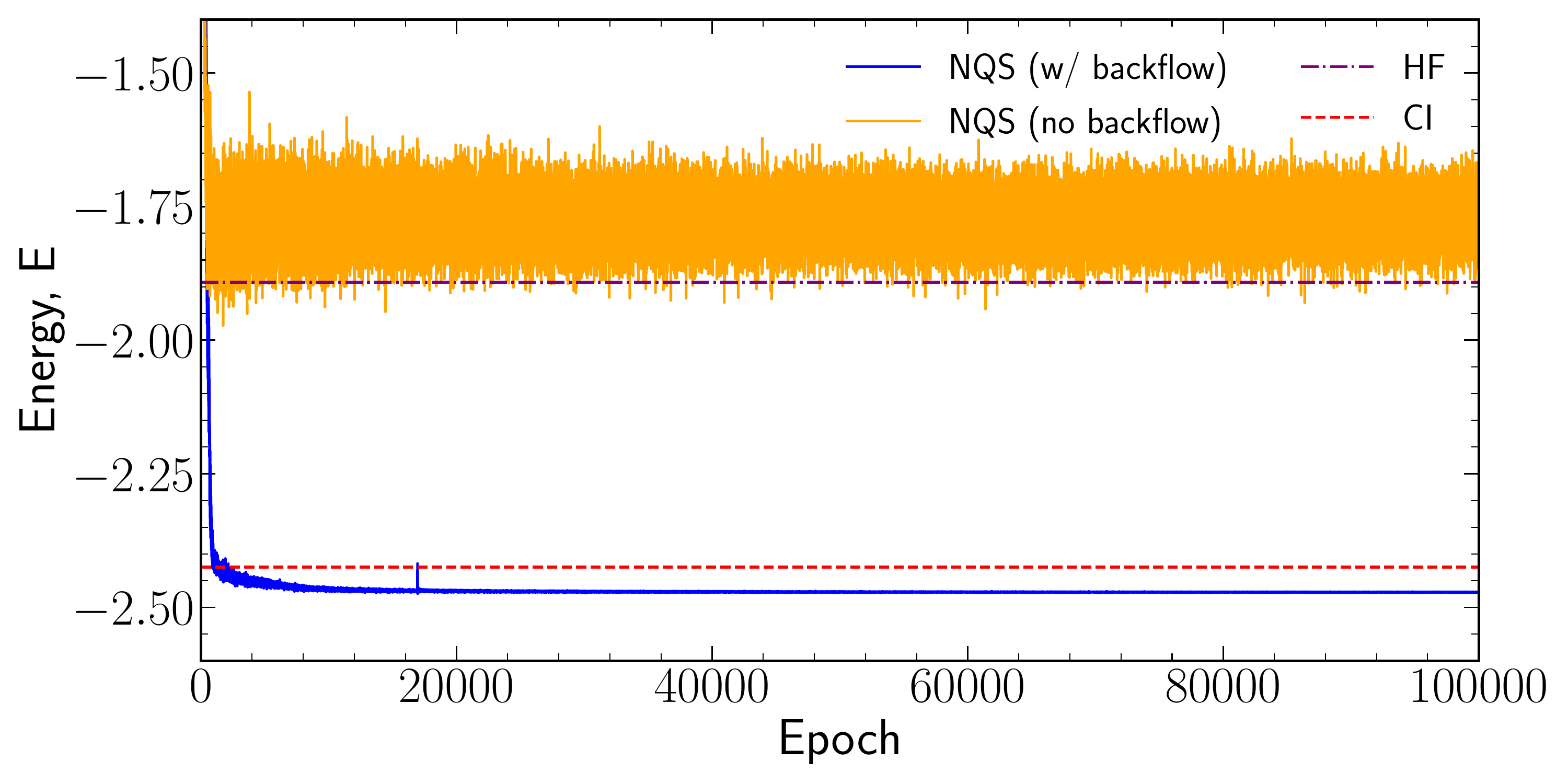}
    \caption{
    Convergence of the expectation value of the energy for the $A = 2$ system with $V_0=-20$ and $\sigma_0=0.5$, as a function of the epoch. The standard NQS \emph{Ansatz} (with backflow) is shown as a blue solid line, and overcomes the exact diagonalisation benchmark (dashed red line). An NQS \emph{Ansatz} without backflow (orange line) converges to the HF solution instead. 
    \label{fig:compare-backflow}
    }
\end{figure}

We perform an insightful numerical experiment to clarify the importance of backflow capabilities in our NQS.
The NQS \emph{Ansatz} can be easily modified to remove any backflow, by eliminating the dependence on $\mu$ in the equivariant layers.
This procedure roughly reduces by a half the number of parameters of the model. Moreover, in a network with a single Slater determinant and without backflow transformation, one  expects the variational process to simply lead to the HF determinant. 
This is exactly what we observe in Fig.~\ref{fig:compare-backflow},
where we show energy minimisation curves for $A=2$ particles
with an interaction strength of $V_0=-20$ and a range $\sigma_0=0.5$. The figure provides results for the standard 
NQS architecture (with backflow, blue solid lines) and for a modified NQS without backflow (orange solid lines). 
The NQS without backflow reaches the Hartree-Fock energy, which is the corresponding optimal energy within a pure single-particle framework. 
It also shows substantial oscillations around the energy minimum. 
In contrast, after incorporating backflow correlations, the NQS can reach (and, in fact, outperform) a direct diagonalization benchmark with minimal oscillations. 
We take this as an indication that the presence of backflow in the \emph{Ansatz}, which is a consequence of permutational equivariance, gives the NQS a remarkable ability to accurately represent beyond mean-field correlation between fermions. 
Furthermore, the statistical fluctations of the energy are  mitigated with the addition of backflow, which indicates that backflow is fundamental in achieving accurate ground-state energies.

\subsection{Direct diagonalisation}
\label{subsec:exact}
We have employed a direct diagonalisation (or
configuration interaction) method
to provide 
a benchmark for the $A$-body NQSs. 
Specifically, we use the HO 
as the single-particle basis.
We express the Hamiltonian of Eq.~(\ref{eq:hamiltonian}) in second quantization,
\begin{align}
\hat{H}= 
\sum_{\alpha=1}^A \epsilon_\alpha \hat{n}_\alpha 
+
\frac{1}{2}
\sum_{\alpha,\beta,\gamma,\delta} V_{\alpha\beta,\delta\gamma} \hat{a}_\alpha^\dagger
\hat{a}_\beta^\dagger \hat{a}_\gamma \hat{a}_\delta,
\end{align}
where 
$V_{\alpha\beta,\delta\gamma}$ 
is the matrix element in the single-particle HO basis, with an analytical expression 
given in Ref.~\cite{mujal2017}.
$\hat{n}_\alpha=\hat{a}_\alpha^\dagger \hat{a}_\alpha$ is the
number operator, and $\hat{a}_\alpha^\dagger$ ($\hat{a}_\alpha$) are the fermionic 
creation (annihilation) operators of the HO state $\alpha$.

In our implementation, we truncate the basis according 
to the many-body noninteracting energy,
which also defines a single-particle
basis truncation. We consider only the first $N_\text{max}=20$ single-particle HO modes 
for all particle numbers. 
This is the maximum number of states allowed by our current resources,
which are limited by the precision in the calculation of the interaction matrix elements. Even with these 
limitations, the results of the repulsive regime are converged. 
In general, the convergence is slower for  attractive 
interactions, as reported  for instance in the zero-range case in 
Ref.~\cite{rojofrancas2023}. 
Having created the many-body basis, we then construct 
the Hamiltonian matrix and diagonalize it using the standard Lanczos algorithm. For details of this method, please refer to Refs.~\cite{plodzien2018, RojoFrancas2020}. 

Our 
calculations are limited by the basis truncation 
$N_\text{max}$, which means that the results 
obtained through diagonalization are not exact and can only provide upper-bound 
energies. Using a larger basis would lead to better results, but the 
many-body basis dimension scales exponentially with the number of particles, making 
it impractical for large systems.

As a consequence of using the HO basis, 
we obtain less error  
in the energy calculation for the repulsive interaction regime than for an attractive interaction. As we shall see below, this can be understood in terms of the spatial rearrangement of the system.  
However, as the number of particles increases even without changing the single-particle 
basis, the accuracy of the results diminishes. In this article, we have only 
considered the ground state, but we stress that this method also provides predictions for excited states.

\subsection{Hartree-Fock}
\label{subsec:hartreefock}

We provide an additional benchmark by looking at the HF ground state, which is a minimal uncorrelated \emph{Ansatz} for the many-body wavefunction. 
We solve the problem in coordinate space, in keeping with the NQS implementation.
This representation is particularly useful in capturing the relatively large changes in shape of the density
distribution, which may otherwise be hard to describe with fixed-basis approaches. 

The HF orbitals $\phi_\alpha(x)$ with 
$\alpha=0, \ldots, A-1$ are fully occupied, with 
occupation numbers $n_\alpha=1$. 
All the remaining states with $\alpha \geq A$ are empty,
$n_\alpha=0$.
The HF orbitals are used to construct a one-body density matrix,
\begin{align}
\rho^{\text{HF}}(x_1',x_1) = \sum_{\alpha=0}^{A-1} \phi^*_\alpha(x_1') \phi_\alpha(x_1),
\label{eq:OBDM_HF}
\end{align}
and the corresponding density profile $n^{\text{HF}}(x)=\rho^{\text{HF}}(x_1=x,x_1'=x)$.
The orbitals are obtained from the HF equations, 
\begin{align}
\left[ - \frac{1}{2} \nabla^2 + \frac{1}{2} x^2 \right] \phi_\alpha(x)
+ \int d \bar x \Sigma(x,\bar x) \phi_\alpha(\bar x)
= \varepsilon^{\text{HF}}_\alpha \phi_\alpha(x) \, .
\label{eq:HF}
\end{align}
In the case of a finite-range interaction, the HF equations are a set of $A$ 
integro-differential equations. The HF self-energy is the sum of a
direct and a (non-local) exchange term,
\begin{align}
\Sigma(x_1',x_1) =& \delta( x_1 - x_1' ) \int d x \, V( x_1-x) n^{\text{HF}}(x) \nonumber \\
&+ V(x_1'-x_1) \rho^{\text{HF}}(x_1',x_1) \, .
\label{eq:HF_meanfield}
\end{align}
The corresponding HF energy is computed from the sum of single-particle energies
and, as a consistency check, it can also be obtained from the direct integral of the mean-field, with 
the associated antisymmetry corrections~\cite{Negele1987}.

We solve this set of self-consistent equations by iteration on a discretized mesh 
of equidistant points. 
The kinetic term is represented as a matrix using the Fourier grid Hamiltonian method~\cite{Marston89}, which works so long as the mesh extends well beyond the support of the wave functions. The mean-field can be computed efficiently with matrix
operations, and Eq.~(\ref{eq:HF}) is reduced to an eigenvalue problem. 
We typically employ $N=200$  grid points extending from $x=-5$ to $5$ for
small systems. For systems with $A>4$, which have larger sizes, the mesh 
limits are extended to $\pm 6$ and we choose $N=240$ points to keep the same mesh spacing.
A computational notebook for the solution of the HF equations 
is available in Ref.~\cite{github_arnau}.

\subsection{Real-space solution for $A=2$}
\label{subsec:real}

In the following section, we use the $A=2$ system to determine which values of interaction strength and range are interesting. We have exploited two more numerical methods  to solve specifically this problem. 

For $A=2$ particles, the Hamiltonian in Eq.~(\ref{eq:hamiltonian2}) is particularly easy to handle. 
Following Refs.~\cite{Busch1998,Koksik2018}, we introduce a center of mass (CoM), 
$R=\frac{1}{\sqrt{2}}(x_1+x_2)$, and relative coordinate, $r=\frac{1}{\sqrt{2}}(x_1-x_2)$. With these, the Hamiltonian $\hat H=\hat H_\text{CM}+\hat H_\text{rel}$ separates into two commuting components,
\begin{align}
\hat H_\text{CM} &= - \frac{1}{2} \nabla_R^2 + \frac{1}{2} R^2 , \\
\hat H_\text{r} &= - \frac{1}{2} \nabla_r^2 + \frac{1}{2} r^2 + \frac{V_0}{\sqrt{ 2 \pi} \sigma_0} e^{ - \frac{ r^2}{\sigma_0^2}} \, .
\end{align}
The center-of-mass component is just an HO. The relative Hamiltonian includes an HO as well as the Gaussian interaction component. We note that the Gaussian form factor in relative coordinates has a different width 
and it is effectively wider than the original interaction. 

The CoM and relative Hamiltonians can be diagonalized separately, providing eigenstates $\varphi_\alpha(R)$ and $\psi_\beta(r)$ with eigenenergies $\epsilon_\alpha$ and $\varepsilon_\beta$, respectively. 
For the relative coordinate, we need to solve the eigenvalue problem
\begin{align}
\left[ - \frac{1}{2} \nabla_r^2 + \frac{1}{2} r^2 + \frac{V_0}{\sqrt{ 2 \pi} \sigma_0} \exp( - \frac{ r^2} {\sigma_0^2} ) \right] \psi_\beta(r) = \varepsilon_\beta \psi_\beta(r) \, .
\label{eq:rel_energy}
\end{align}
Because of the Pauli principle, the only acceptable solutions are those that are antisymmetric in the relative coordinate, $\psi_\beta(-r)=-\psi_\beta(r)$.  
These correspond to odd values of $\beta$, and hence the ground state will necessarily correspond to $\beta=1$ rather than lowest-energy, space-symmetric $\beta=0$ state. To solve this problem numerically, we discretize the relative coordinate in an evenly spaced mesh. As in the HF case, the kinetic term is discretized as a matrix using the Fourier grid Hamiltonian method~\cite{Marston89}. We employ $N=200$ grid points extending from $r=-5$ to $5$. A computational notebook for this problem is also available in Ref.~\cite{github_arnau}.

If a numerical solution to $\psi_\beta$ is available, the total two-body wavefunction is the product,
\begin{align}
\Psi_{\alpha,\beta}(x_1,x_2) = \varphi_\alpha \left( \frac{1}{\sqrt{2}}(x_1+x_2) \right) 
\times \psi_\beta \left( \frac{1}{\sqrt{2}}(x_1-x_2) \right) .
\end{align}
The total energy of the ground state
is given by 
$E = \epsilon_0 + \varepsilon_1 = \frac{1}{2} + \varepsilon_1$. We only discuss the ground state of the system, but note that this method can  also provide the rest of the spectrum for $A=2$. 

\subsection{Perturbation theory for $A=2$}

Alternatively, one can solve for the energy $\varepsilon_\beta$ in Eq.~(\ref{eq:rel_energy}) using standard perturbation theory tools. We take the HO as a reference state, so that $\psi_\alpha^{(0)}=\varphi_\alpha$ and $\varepsilon_\alpha^{(0)}=\epsilon_\alpha$, 
where $\varphi_\alpha$ and $\epsilon_\alpha$ are HO eigenstates and eigenvalues. The Gaussian interaction term is then treated as a perturbation. We employ known analytical expressions for the matrix elements  
$V_{\alpha \beta}=\frac{V_0}{\sqrt{ 2 \pi} \sigma_0}  
\Braket{ \varphi_\alpha | e^{ - \frac{ r^2} {\sigma_0^2} } | \varphi_\beta}$
\cite{Earl2008}. 
The zero-order results are just the eigenvalues of the noninteracting confined two-particle system, Eq.~(\ref{eq:E_A_noninteracting}),
which in the ground state yields $E^{(0)}=2$.
Noting that antisymmetry constraints force $\alpha$ and $\beta$ to take odd values, we find that the first-order perturbation theory (PT1) expression for the total energy is, 
\begin{align}
    E^{(1)}= E^{(0)} + V_{11} 
    = 2 + \frac{V_0}{\sqrt{2 \pi}} \frac{\sigma_0^2}{(1+\sigma_0^2)^{3/2}}.
    \label{eq:PT1} 
\end{align}
As expected, this expression is linear in the perturbation strength $V_0$.
The slope of the energy dependence on $V_0$ is dictated by $\sigma_0$. For
small values of $\sigma_0$, the slope grows quadratically. 
In other words, the departure from the noninteracting case is quadratic in
$\sigma_0$. For large values of 
$\sigma_0$, in contrast, the slope decreases like $1/\sigma_0$. This is to be expected since,
by construction, the parametrization of our interaction term has such a $1/\sigma_0$ dependence.  

Second (PT2) and third (PT3) order perturbation theory  results can be readily obtained from the corresponding matrix elements $V_{\alpha \beta}$. We use up to eight additional states in the intermediate sums, which are already converged for all practical purposes. 
We provide a computational notebook in Ref.~\cite{github_arnau}. 
As we shall see below, these PT estimates allow us to find regions of parameter space where nonperturbative effects are particularly important.

\section{Results}
\label{sec:results}

In this section, we show the results obtained for spinless fermionic systems from $A=2$ to $A=6$ with different methods. We start with a discussion of
the two-body case in the first subsection. This allows us to perform an exploration of the dependence of the results 
in $V_0$ and $\sigma_0$. Results for $A>2$ are discussed in the following subsection, Sec.~\ref{sec:few-body}.

\subsection{Two-body sector}

The two-body case in our Gaussian toy model is already relatively complicated. The solution for a zero-range interaction
is well known~\cite{Busch1998}, and semianalytic solutions for specific finite-range interactions are also
available~\cite{Koksik2018}.

\subsubsection{Energy}

The dependence on $V_0$ and $\sigma_0$ of the energy of the $A=2$ mimics that 
of heavier systems.  We are particularly 
interested in finding regions of nonperturbative behaviour, to test the performance of the NQS \emph{Ansatz} in the most challenging scenarios.
We start by looking at the dependence on the range of the interaction, $\sigma_0$. We note that our pursuit here is mostly theoretical and that the values of interaction range and strength that we explore may not be directly accessible by near-term experiments. 

\begin{figure}[t]
\includegraphics[width=\linewidth]{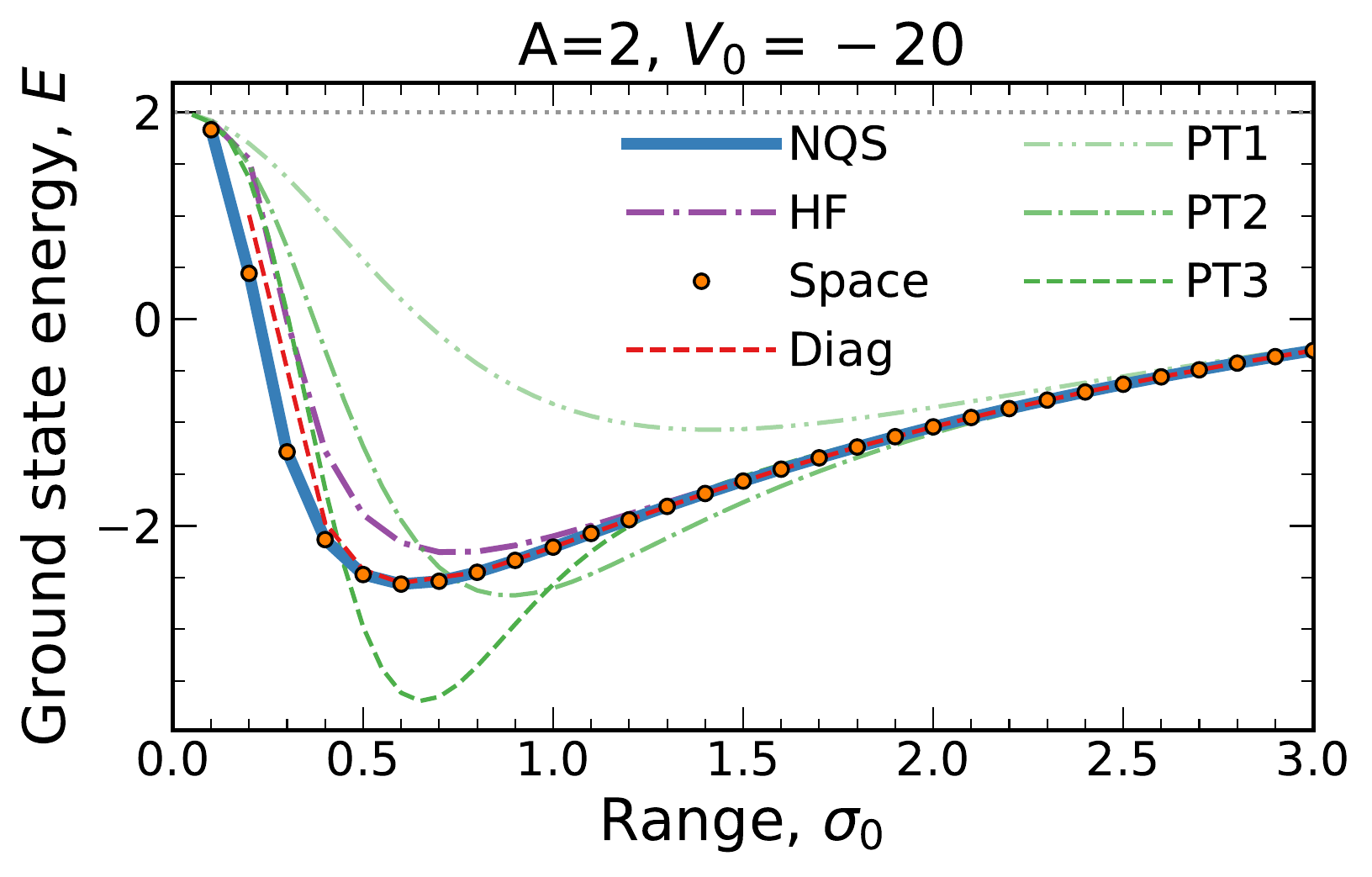}
\caption{
The ground state energy of the $A=2$ system as a function of the range, $\sigma_0$, for a large and attractive interaction strength, $V_0=-20$. The NQS \emph{Ansatz} energy for $H=64$, $L=2$, and $D=1$ is shown as a solid line. The orange circles 
represent the  real-space solution. The short-dash-dotted purple line is the Hartree-Fock solution. 
The double-dotted-dashed, dotted-dashed, and dashed green lines show the perturbation theory results to first (PT1), second (PT2), and third orders (PT3), respectively. The horizontal dotted line is the noninteracting energy baseline. }
\label{fig:energy_A2_range_V0=-20}
\end{figure}

We start by considering a very attractive interaction, with $V_0=-20$. 
The results for the ground-state energy are 
shown in Fig.~\ref{fig:energy_A2_range_V0=-20}. In this plot, we show the NQS \emph{Ansatz} with a solid line, which includes the statistical VMC uncertainty that is typically smaller than $10^{-5}$ in the same units. We also show the exact diagonalization (dashed line), 
HF (short-dashed-dotted purple line) and the real-space (filled circles) solutions.
To analyze how perturbative
the results are, we also display the PT1, PT2 and PT3 predictions for the energy with different linestyles. 

For $A=2$ particles, the
noninteracting case corresponds to $E=2$, shown in the figure as a horizontal dotted line. 
As expected, all the methods agree with this value as the range
tends to zero, $\sigma_0 \to 0$. As a function of the range,  all energy predictions subsequently decrease, reach a model-dependent minimum value and eventually increase to reach
an asymptotic $\sigma_0^{-1}$ behaviour. 

The comparison between different methods provides an insight on the complexity of the problem. First, we note that the NQS solution agrees
perfectly well with the real-space solution along a wide range of values. Second, the minimum of energy as a function of $\sigma_0$ lies around 
$\sigma_0 \approx 0.5$ and yields $E \approx -2$ for this particular value of $V_0$. We conclude that the network is performing well across a wide range of values of $\sigma_0$.

The NQS outperforms variationally the exact diagonalization and HF solutions across a wide range of values. In particular, the HF solution has a somewhat shallower minimum
at a larger value of $\sigma_0$. Note that the HF 
prediction is the optimal solution
for a Slater determinant formed of single-particle orbitals, $\phi_\alpha(x)$~\cite{martin_reining_ceperley_2016,Blaizot1986,Gross1991}.
The HF single-particle orbitals depend on a single position. 
The NQS outperforms this solution by exploiting backflow correlation in the generalized orbitals of the GSM, as explained in Subsection~\ref{subseq:backflow}. 
The NQS prediction is also better than the exact diagonalization results at 
very small values of $\sigma_0$, where the truncated basis may have difficulties capturing
the very narrow structures appearing in the $\sigma_0 \to 0$ limit.

\begin{figure}[t]
\includegraphics[width=\linewidth]{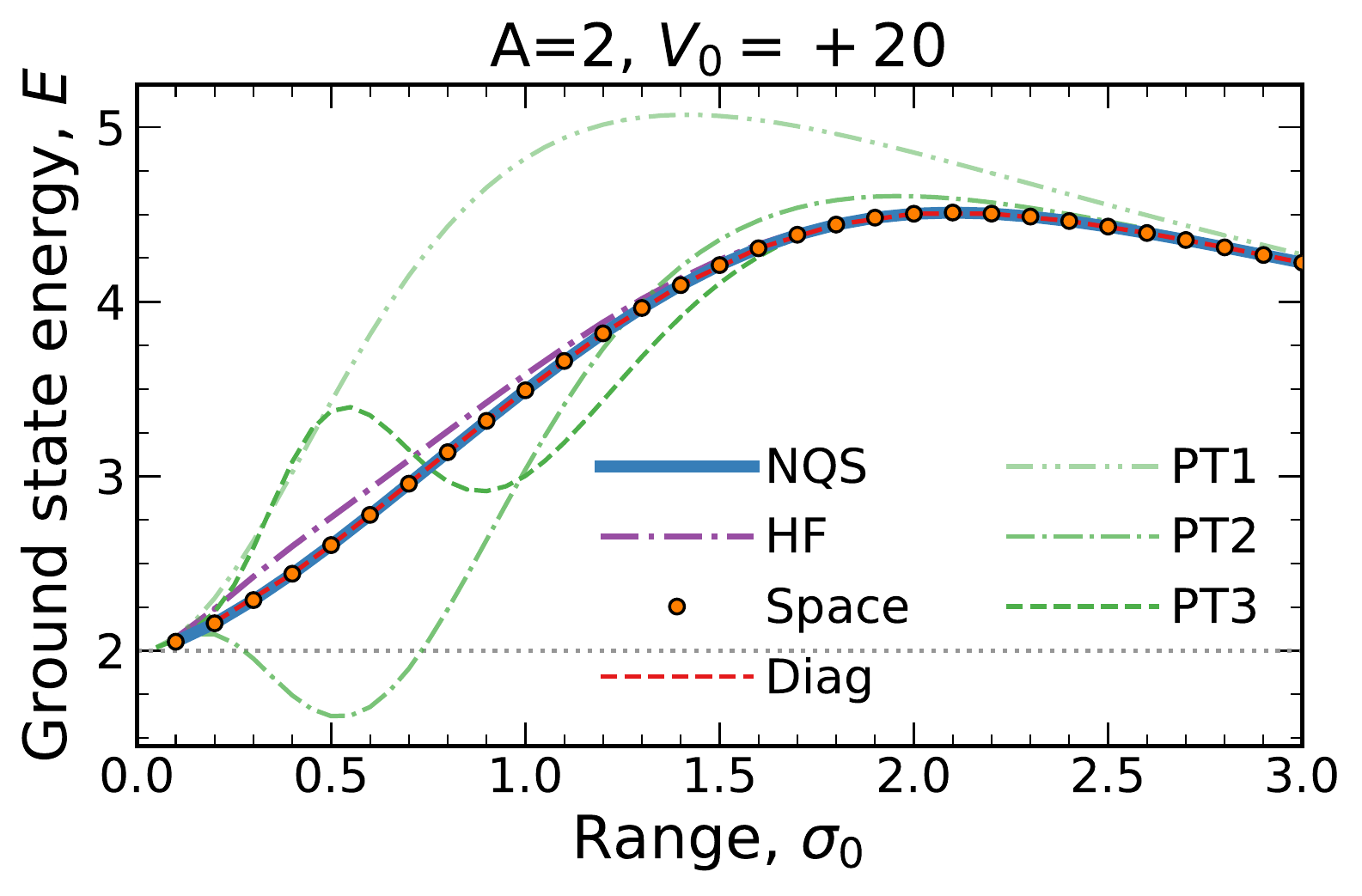}
\caption{
The same as Fig.~\ref{fig:energy_A2_range_V0=-20} but for a large and repulsive interaction strength, $V_0=+20$.}
\label{fig:energy_A2_range_V0=+20}
\end{figure}

The convergence of the perturbation theory results provides an indication of the ``perturbativeness'' of the two-body
problem.  The PT1 (dashed-double-dotted line) result of Eq.~(\ref{eq:PT1}) is far too repulsive across all values of the range. 
It only gets close to the NQS and real-space results well beyond its shallow minimum, which happens at $\sigma_0 = \sqrt{2}$. 
The PT2 results for small values of $\sigma_0$
are substantially closer to the NQS ones, but they cannot reproduce
the correct dependence with $\sigma_0$ below about $\sigma_0 \approx 2$. In contrast to the 
PT1 prediction, the PT2 result is more attractive than the NQS prediction beyond about $\sigma_0 \approx 0.7$. 
Something similar happens with PT3, which is well below the NQS solution from $\sigma_0 \approx 0.5$ onwards. 
We find that these third-order results are still rather far from the true values and, in particular, the convergence pattern is relatively erratic for $\sigma \in [0.5, 0.9]$. This challenging region in parameter space may be a good test bed for the NQS solution.

We confirm that this region of ranges is particularly nonperturbative by looking at a very repulsive case. 
We show in Fig.~\ref{fig:energy_A2_range_V0=+20} the predictions for the ground state energy of the $A=2$ system for a value of $V_0=+20$. 
The shape of the energy predictions here is substantially different. The NQS energy departs 
from the noninteracting
baseline at $\sigma_0=0$, and increases to a maximum of around $E \approx 4$ at $\sigma_0=2$. Just as in the attractive case, large values of the range lead to relatively perturbative results, in the sense that 
PT1, PT2 and PT3 predictions agree with the exact benchmarks. In contrast, for values of the range below $\sigma_0=1$, the PT predictions follow a complex pattern. In particular, the PT2 and PT3 results show anomalous 
oscillations around $\sigma_0=0.5$. All in all, we conclude that for values of interaction strength of the order
of $V_0 \approx \pm 20$, nonperturbative behavior occurs for interaction ranges of the order of 
$\sigma_0=0.5$. While we do not show them for brevity here, we stress that the dependence 
of the NQS and HF energies on the range $\sigma_0$ for systems with $A>2$ has a behavior very similar to those shown in Figs.~\ref{fig:energy_A2_range_V0=-20} and \ref{fig:energy_A2_range_V0=+20}. The interested reader can find this
information in Ref.~\cite{Keeble_thesis}.

\begin{figure}[t]
\includegraphics[width=\linewidth]{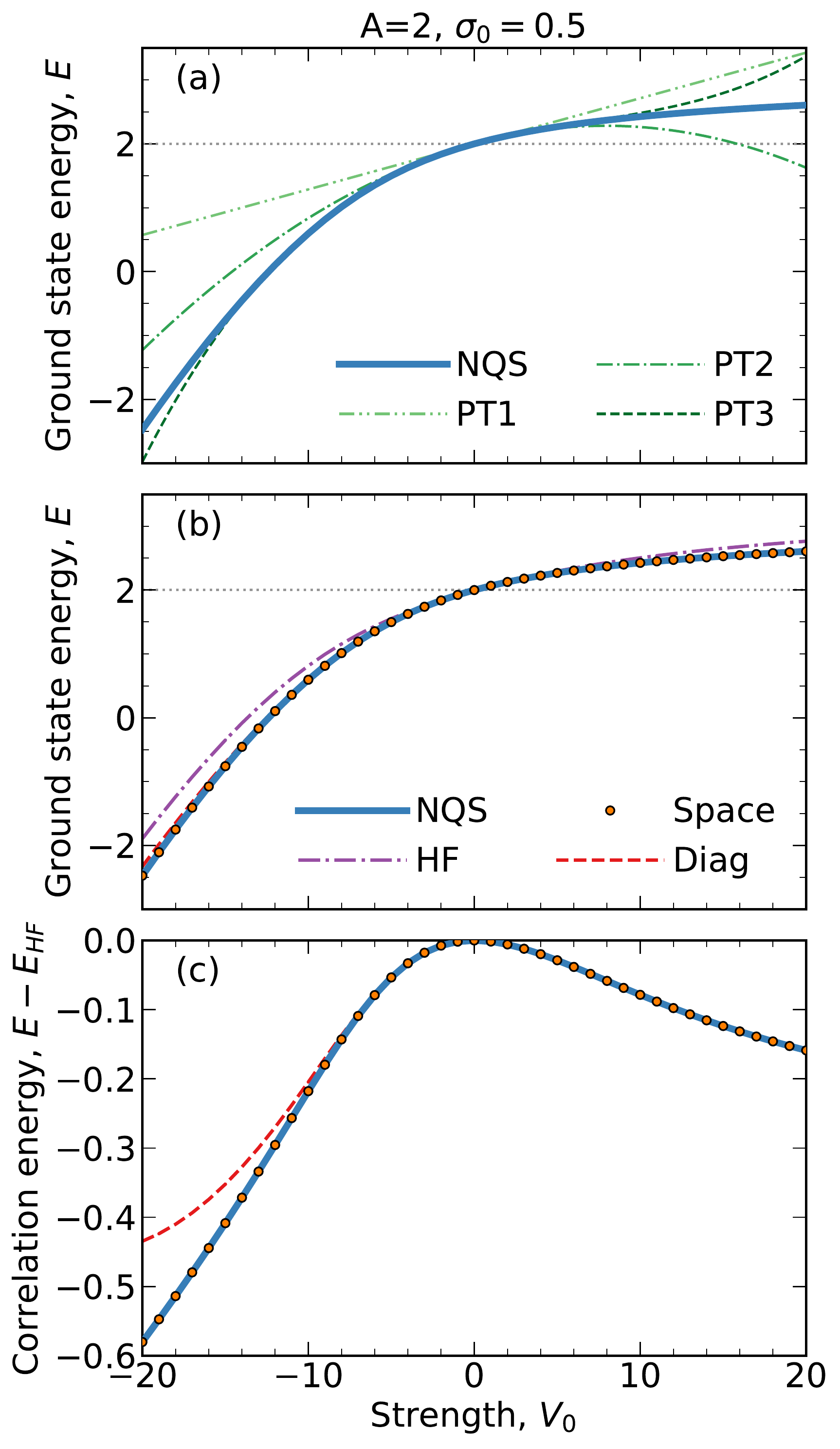}
\caption{Energy of the $A=2$ system as a function of interaction strength, $V_{0}$, for a fixed range, $\sigma_0=0.5$. 
Panel (a): comparison of the NQS ground state energy (solid line) to perturbation theory results.
Panel (b): comparison of the NQS ground state energy (solid line) to the real-space solution of the
$A=2$ system (filled circles); direct diagonalisation results (dashed red line); and the HF approximation (short-dashed-dotted purple line).
Panel (c): NQS, real-space solution and direct diagonalisation 
predictions for the correlation energy, $E_c=E-E_\textrm{HF}$.
}
\label{fig:energy_A2_strength}
\end{figure}

\begin{figure*}[t]
\includegraphics[width=\linewidth]{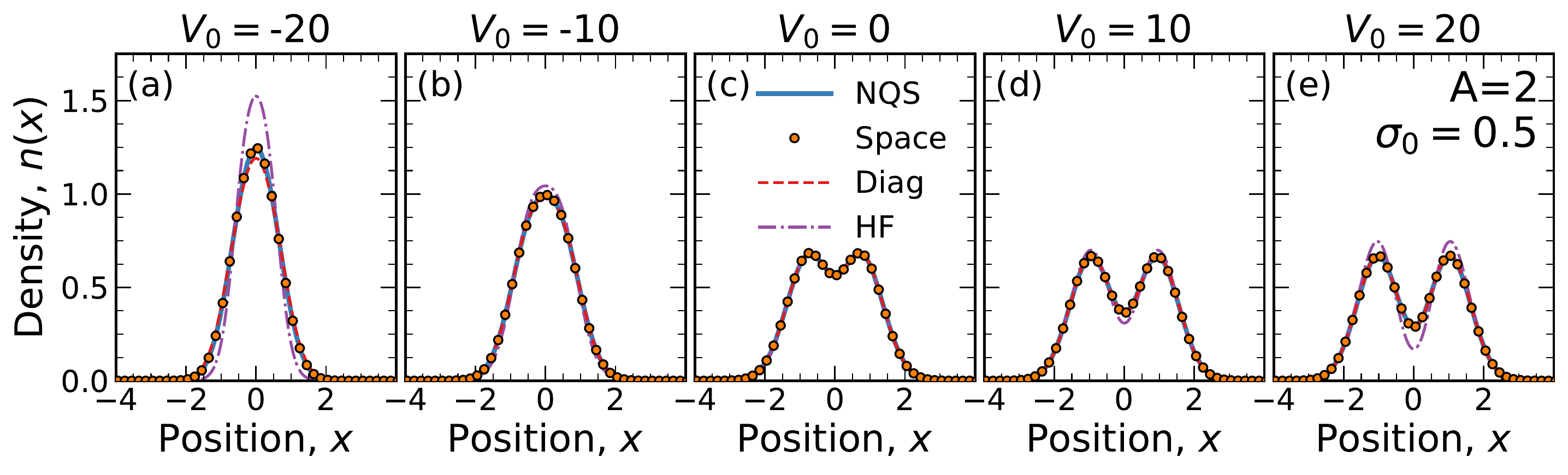}
\caption{The density profile $n(x)$ of the $A=2$ system as a function of position $x$.  Panels (a) 
to (e) show results for values of $V_0$ from $-20$ to $20$ in steps of $10$.
See Fig.~\ref{fig:energy_A2_strength} for an explanation of the legend.}
\label{fig:density_A2}
\end{figure*}

The analysis of the $\sigma_0$ dependence of our results
leads us to conclude that a value of $\sigma_0=0.5$ 
is 
well within the nonperturbative regime. We now explore
the dependence of the ground-state energy on the interaction strength $V_0$ to analyze
the performance of the NQS \emph{Ansatz}. The results for $A=2$ are shown in the three 
panels of Fig.~\ref{fig:energy_A2_strength}. 
Figure~\ref{fig:energy_A2_strength}(a) shows a comparison between the NQS ground state energy (solid lines)
and the different PT results.
This provides an idea of the perturbative nature of the system
as a function of $V_0$, rather than $\sigma_0$. The PT1
prediction (dashed-double-dotted line) of Eq.~(\ref{eq:PT1}) is, as expected, linear in $V_0$ and captures
the full dependence on $V_0$ only for small values of $V_0$, $| V_0 | < 4$. PT2 
results (dashed-dotted line) are valid in a wider strength range, but overpredict
(underpredict) the true ground-state energy at large negative (positive) values
of $V_0$. PT3 results (short-dashed line) work well within a range $-15 < V_0 < 10$, but fail
beyond this range. Again,
this analysis 
indicates that the most attractive and
repulsive values of $V_0$ are in a nonperturbative regime.

We compare the NQS results to more realistic benchmarks in Fig.~\ref{fig:energy_A2_strength}(b). Overall, we find an excellent 
agreement between the real-space solution and the NQS prediction. The direct 
diagonalization result also provides a good energy prediction, although
minor deviations with respect to the NQS are observed at large negative
values of the coupling. The NQS provides better variational
results than the HF approximation thanks to backflow
correlations. 

We now comment on the overall dependence of the energy on $V_0$. In the noninteracting limit $V_0=0$, all methods agree exactly with the baseline value, $E=2$. In the repulsive side, as the interaction strength increases in magnitude, the energy generally increases above the baseline, and seems to saturate to a value that lies close to $E \approx 2.5$. 
On the attractive side,
in contrast, the energy decreases relatively rapidly. At around $V_{0} \approx -10$, the system energy becomes attractive. The total energy then decreases
in absolute value as $V_0$ becomes more and more attractive. 
It is immediately clear from the energetics of the system
that the attractive and the repulsive regimes are very different from each other. 

This difference becomes even more evident in Fig.~\ref{fig:energy_A2_strength}(c), which shows the
correlation energy $E_c$ of the system. We follow textbook conventions~\cite{martin_reining_ceperley_2016} and define $E_c$ as the difference in 
ground-state energy of a given method and the HF prediction, $E_c=E-E_\text{HF}$. 
We note that this may not be a very insightful quantity at the two-body level, since the HF prediction is not expected
to work well in the $A=2$ sector. 

The correlation energy departs from zero at $V_0=0$. 
In the repulsive side, $E_c$
slowly increases in absolute magnitude. At $V_0=20$, we find $E_c \approx -0.15$, which
is about $6 \%$ of the total energy. In contrast, in the attractive 
regime, the NQS prediction is $E_c \approx -0.6$, which is a substantial contribution
to the total energy of $E \approx -2$ at that same value. There is no sign of saturation
of the correlation energy in either side of $V_0$. We also note that the direct
diagonalization result departs from both the real-space and the NQS prediction for 
values of $V_0 < -10$. This discrepancy is due to the truncation of the
model space, as we shall see next.

\subsubsection{Density profile}

To further understand the origin of the correlations of the system, we
look at the density profile of the system, $n(x)$. 
Figure~\ref{fig:density_A2} shows the density profile obtained with 
the different many-body approaches for five different values of 
$V_0$, ranging from $-20$ to $20$ in steps of $10$. 
Figure~\ref{fig:density_A2}(c) shows the noninteracting result, $V_0=0$. Here, all methods agree, as 
expected. The density profile has a well-understood dip at the center, due to the 
interplay between the $\alpha=0$ and $\alpha=1$ HO orbitals. 
These density modulations
are typical in fermion systems~\cite{Koksik2018}. 
Figures~\ref{fig:density_A2}(d) and ~\ref{fig:density_A2}(e) show the same densities
in the repulsive regime, 
at interaction values of $V_0=10$ and $20$, respectively. We find that the 
central dip of the density decreases, while the overall size of the density 
profile is relatively constant. We find a very clear agreement between the 
real space solution of the $A=2$ problem, the direct diagonalization method  
and the NQS prediction. In contrast, the HF result seems to overestimate 
the fermionic structures, with a lower dip and higher density maxima. In
other words, the HF orbitals appear to be more localized than their correlated
counterparts. A similar difference in the localization patterns of HF and 
fully correlated predictions was observed in the 
3D Wigner crystal of an electron gas~\cite{Drummond2004}. 

Physically, the picture that arises in the repulsive regime is akin to that of localization or (Wigner) 
crystallization~\cite{wigner1934interaction,Jauregui1993,Szafran2004,Shapir2019,DinhDuy2020,Ziani2020,Luo2023}. As the 
interparticle interaction strength increases, the system minimizes the 
interaction energy by keeping particles beyond the interaction range. Doing so, however, may come at the price of increasing the potential energy if particles
were moved away from the center of the HO well. Instead, the system prefers to
localize orbitals more strongly in a periodic structure, reducing the inter-particle density in the intermediate throughs. 

In contrast, the density profile in the attractive regime, shown in Figs.~\ref{fig:density_A2}(a) and ~\ref{fig:density_A2}(b), is characterized by a relatively 
featureless structure. The fermionic dip disappears around 
$V_0 \approx -10$ and, for more attractive strength values, the density profile is a simple 
peak that narrows down and increases in height as $V_0$ becomes 
more negative. 
In fact, a longstanding prediction in one-dimensional systems suggests that 
spinless fermionic systems with strongly attractive 
interactions should behave like noninteracting bosonic 
systems~\cite{Girardeau2003,Granger2004,Girardeau2004,Girardeau2004b,Valiente2020,Valiente2021,Morera2022,Koksik2018,Koksik2020}.
The density profile for the $A=2$ system clearly supports this
hypothesis. It is worth stressing that the NQS grasps the 
bosonization transition without any further adjustments. 
We also stress that our results are qualitatively similar to the semi-analytical 
solution of Ref.~\cite{Koksik2018}, where a soft-core, square-well interaction was used. 

Overall, the density profile of the system is extremely useful because it provides 
direct evidence of two very different phases of the toy model. 
The attractive regime shows signs of a bosonic nature in the density of the
system, a prediction that bodes well with previous studies employing odd-symmetry interaction potentials~\cite{Girardeau2003,Granger2004,Girardeau2004,Girardeau2004b,Valiente2020,Valiente2021,Morera2022,Koksik2020}. 
Very few predictions exist for parity-even potentials~\cite{Koksik2018,Schilling2013}, but the agreement between our different benchmarks in the bosonic phase indicates that the picture is relatively robust. 
A previous theoretical study employing
an even-parity soft-core potential in the $A=2$ case~\cite{Koksik2018} identified similar features in the repulsive phase. There, particles tend
to localize beyond the interaction range, to move away from the repulsive
core and thus minimize the energy. In fact, for repulsive
interactions, the same study indicates that the localized phase is identical for both fermions and bosons~\cite{Koksik2018}. 
All in all, the existing evidence at the two-body level for an even-parity potential indicates that there are signs of fermion-boson duality both in the attractive and the repulsive regimes.

\subsubsection{Occupation numbers}

The density profile $n(x)$ is just a component of the OBDM. We can access the OBDM with the  benchmark methods discussed so far. 
With NQSs, we calculate $\rho(x_1',x_1)$ stochastically using the ghost-particle method of Ref.~\cite{mcmillan1965ground}. 
For the real-space solution, we perform the integral in
Eq.~(\ref{eq:OBDM}) over the uniform mesh grid.
The solution of Eq.~(\ref{eq:occ_numbers}) can then be easily obtained
as a matrix diagonalization problem in a spatially 
uniform grid. 
With the direct diagonalization method, we calculate
$\rho$ in the second-quantization formalism, with the definition
\begin{equation}
    \rho_{\alpha \beta}=
    \langle \Psi| \hat a_\alpha^\dagger \hat a_\beta |\Psi\rangle\,,
\end{equation}
where $|\Psi\rangle$ is the precomputed ground-state wave function and $\rho_{\alpha \beta}$ are the matrix elements of the OBDM, $\rho$. Finally, we diagonalize $\rho$ to obtain the eigenvalues $n_\alpha$ in the many-body basis.

\begin{figure}[t]
\includegraphics[width=\linewidth]{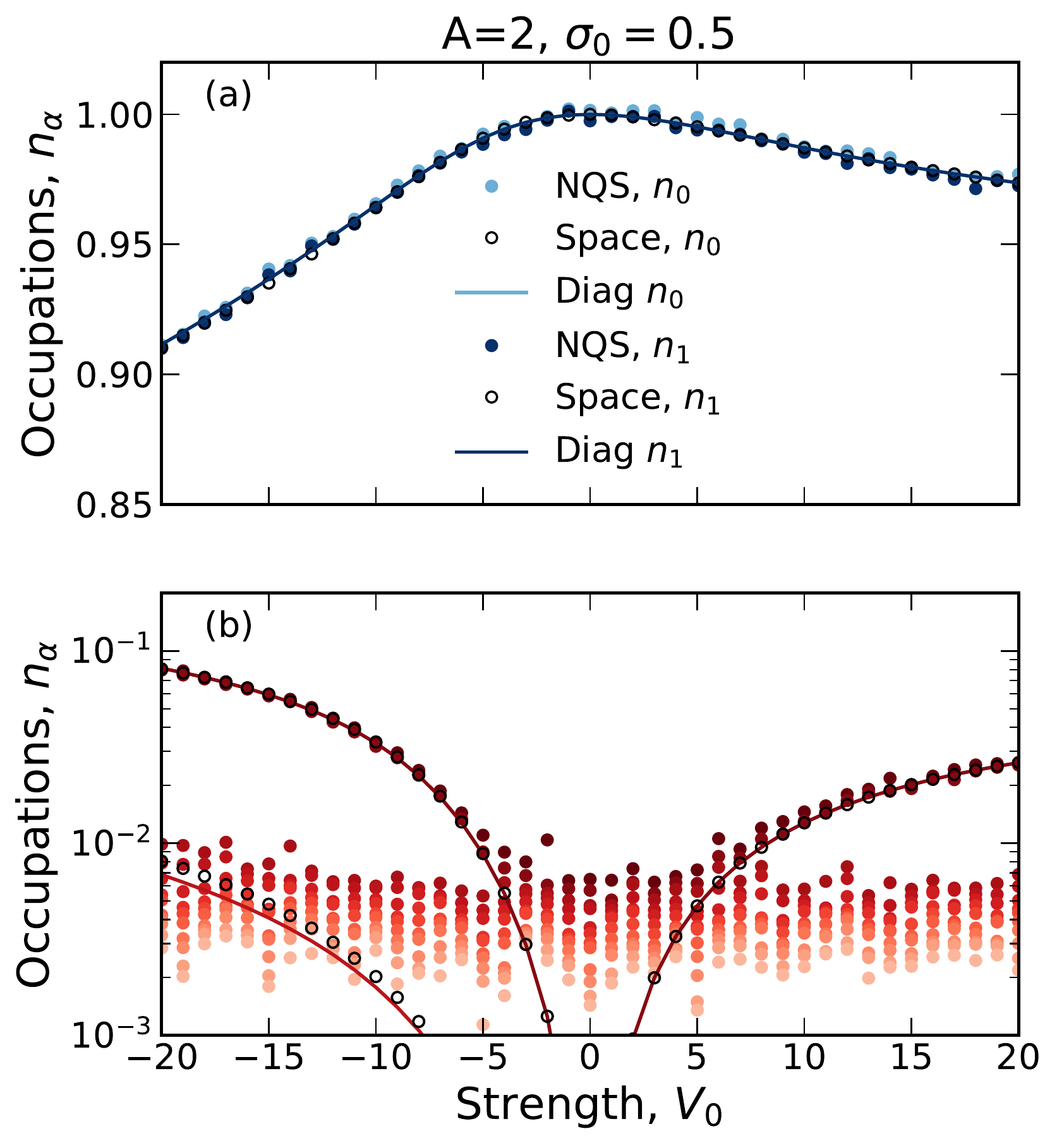}
\caption{Occupation probabilities for the $A=2$ system as a function of interaction strength for a fixed range $\sigma_0=0.5$. 
Panel (a) shows the occupations for the two hole states below the Fermi surface. 
Panel (b) shows (in logarithmic scale) the occupations of particle states. 
We compare the NQS results (solid circles) to real space results (empty 
circles). Solid lines correspond to the direct diagonalisation results. 
}
\label{fig:occupation_A2}
\end{figure}

Figure~\ref{fig:occupation_A2}(a) shows the evolution of the occupation numbers
for the $A=2$ system as a function of interaction strength. We benchmark the NQS results 
(solid circles) to the real-space solution (empty circles) and the direct
diagonalization approach (solid lines). 
We find an excellent agreement between
all three methods. 
We note that we do not include a comparison to the HF benchmark, which provides a relatively good description of the energy of the system. For occupation numbers, however, the HF approach is limited to either fully occupied or fully unoccupied states and it cannot capture deviations in the occupation numbers from the noninteracting case.

For $A=2$, there is a double degeneracy in the natural orbital occupations, 
which the NQS can handle seamlessly. The hole occupation values 
$n_{\alpha < A}$ depart from $1$ as the absolute value of $V_0$ increases. We find
that the rate of change is different depending on whether we are on the
attractive or the repulsive side, a result that mimics the correlation energy displayed
in Fig.~\ref{fig:energy_A2_strength}(c).
In the repulsive side, for $V_0=20$, the occupation is about $\approx 96 \%$,
indicating a relatively uncorrelated system. 
The tendency to deviate from $1$ as $V_0$ increases, however, indicates that 
even in the crystalline phase, particles are not entirely described by locally confined
individual orbitals~\cite{Koksik2018}.
In contrast, in the attractive case, for $V_0 \approx -20$ the
occupation number is $n_\alpha \approx 0.9$. This suggests a stronger role of correlations
in the bosonized phase.

Figure~\ref{fig:occupation_A2}(b) shows a similar figure for ``unoccupied'' orbitals, $n_{\alpha \geq A}$. The plot is
given in logarithmic scale and includes the first $14$ orbitals in the NQS diagonalization. Due to the stochastic nature of the OBDM estimation in the NQS approach,
there is a prominent statistical noise floor of the order of $\approx 10^{-2}$. 
While this limits some of the conclusions that can be drawn, we stress that the NQS 
many-body wave function is comparable (if not better) variationally than the other estimates. The values of $n_\alpha$ above the statistical threshold should therefore provide a more faithful representation of the true occupation numbers.

Just as in Fig.\ref{fig:occupation_A2}(a), we find that the repulsive and attractive sides are qualitatively different. 
On the repulsive regime, we only see one doubly-degenerate particle state with an occupation larger than $10^{-3}$,
that reaches values close to $n_\alpha \approx 0.02-0.03$ for $V_0=20$. 
In the attractive regime, in contrast, we find two doubly-degenerate states with occupations above $10^{-3}$. The more populated states here reach values $n_\alpha \approx 0.08$ for $V_0 = -20$.
The second doubly-degenerate state is observed as a prediction of the exact 
diagonalization and real-space solutions.
It has an occupation which is roughly $10$ times smaller than the most occupied state. 
The NQS cannot resolve this occupation, as it lies within 
its statistical floor. Overall, however, the picture reinforces the idea that correlations have a very different nature on the attractive and the repulsive side. 
\emph{A priori}, it appears that the bosonized phase
has a more complex structure, admixing more single-particle modes than the corresponding 
crystalline phase. 

\subsection{Few-body sector}
\label{sec:few-body}
We now turn our discussion to the results obtained in the few-body sector, for  systems 
with $A=3$ up to $A=6$ particles. 
We continue to benchmark NQS results against 
the direct diagonalization approach as well as the HF method. The real-space solution is not available
for $A>2$. 
We advance that several of the physics conclusions we 
drew on the two-body system remain the same 
in the few-body domain. 

\begin{figure}
\includegraphics[width=0.8\linewidth]{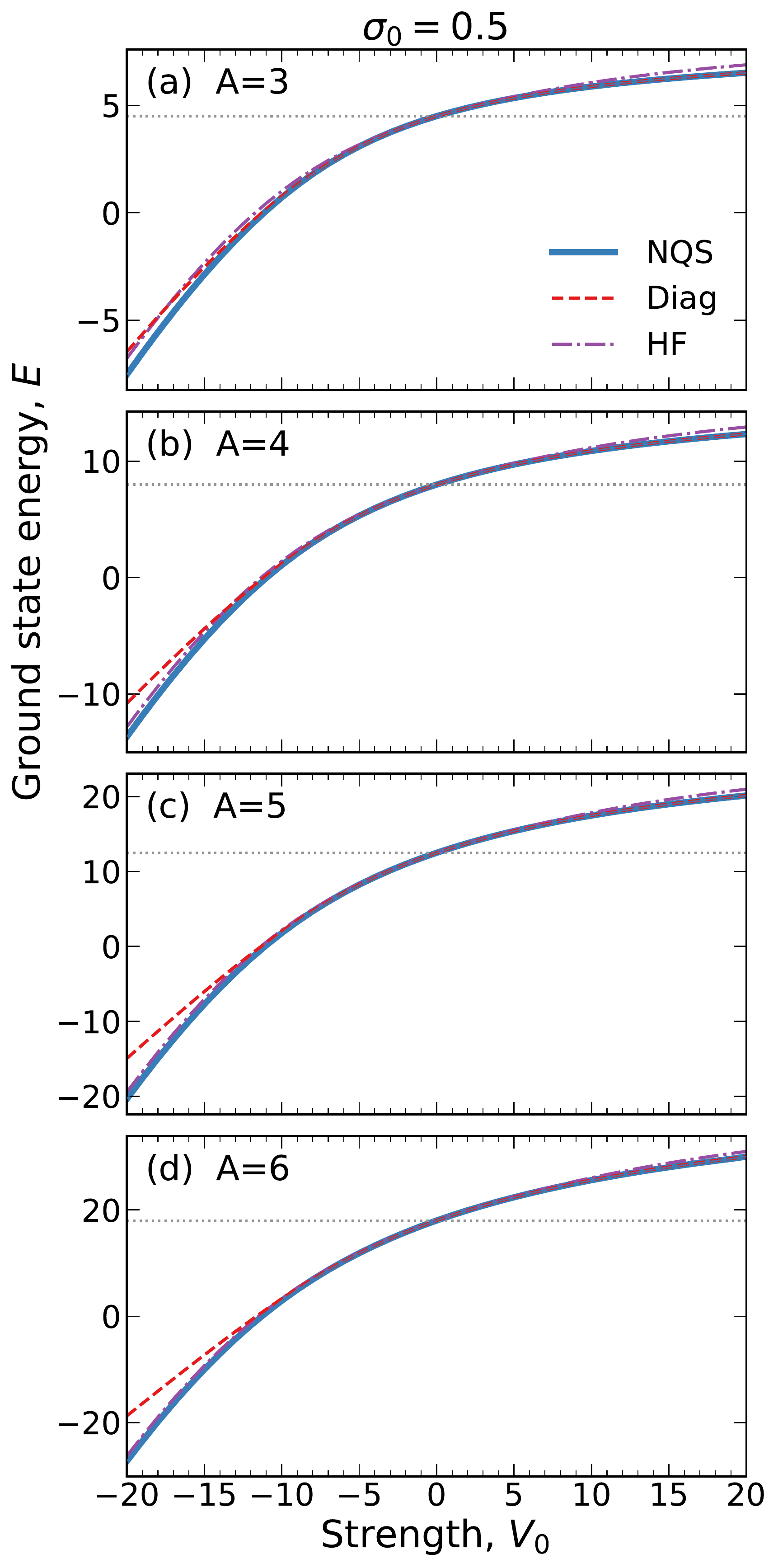}
\caption{
The ground-state energy of the $A$-particle system as a function of interaction strength, $V_{0}$, for a fixed range, $\sigma_0=0.5$. 
Panels (a) to (d) correspond to systems of $A=3$ to $A=6$ particles. 
See Fig.~\ref{fig:energy_A2_strength} for an explanation of the legend.}
\label{fig:energies}
\end{figure}

\subsubsection{Energy}

We start by considering the energy of the system. Figures~\ref{fig:energies}(a)-\ref{fig:energies}(d) show the energy for 
$A=3$ to $6$ particles as a function of $V_0$. 
We focus on the case with a range value of $\sigma_0=0.5$. 
The qualitative behavior of the energy in all these systems is very similar
to that in the $A=2$ case. On the repulsive side, the system energy
increases slowly and tends to plateau at a value which is about 
$1.25-1.6$ times the noninteracting value. In this repulsive range, 
a power-law scaling with the particle number
$E \approx A^{2.2}$ approximates the complete results well. 
The saturation of the energy as $V_0$ becomes very repulsive bodes well
with the physical picture of a crystal. This picture is further
reinforced by the density profiles, which we present later. 
In the attractive regime, in contrast, the energy per particle decreases
much more steeply as a function of $V_0$, with no signs of saturation 
as $V_0$ becomes more negative. 

\begin{figure}[t]
\includegraphics[width=0.8\linewidth]{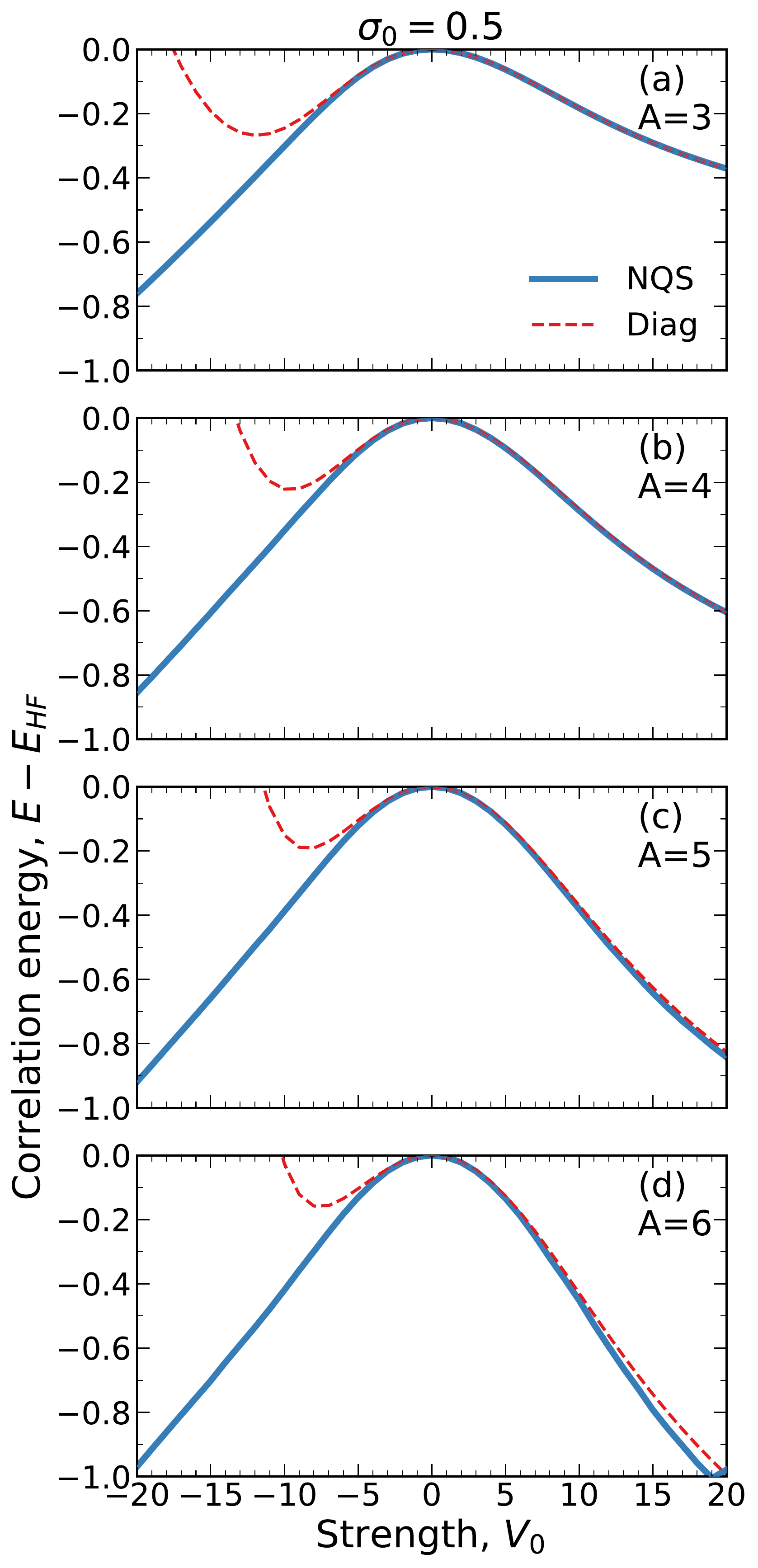}
\caption{
The correlation energy of the $A$-particle system as a function of interaction strength, $V_0$, for a fixed range, $\sigma_0=0.5$. 
Panels (a) to (d) correspond to systems of $A=3$ to $A=6$ particles. 
See Fig.~\ref{fig:energy_A2_strength} for an explanation of the legend.}
\label{fig:correlation_energies}
\end{figure}

\begin{figure*}
\includegraphics[width=0.6\linewidth]{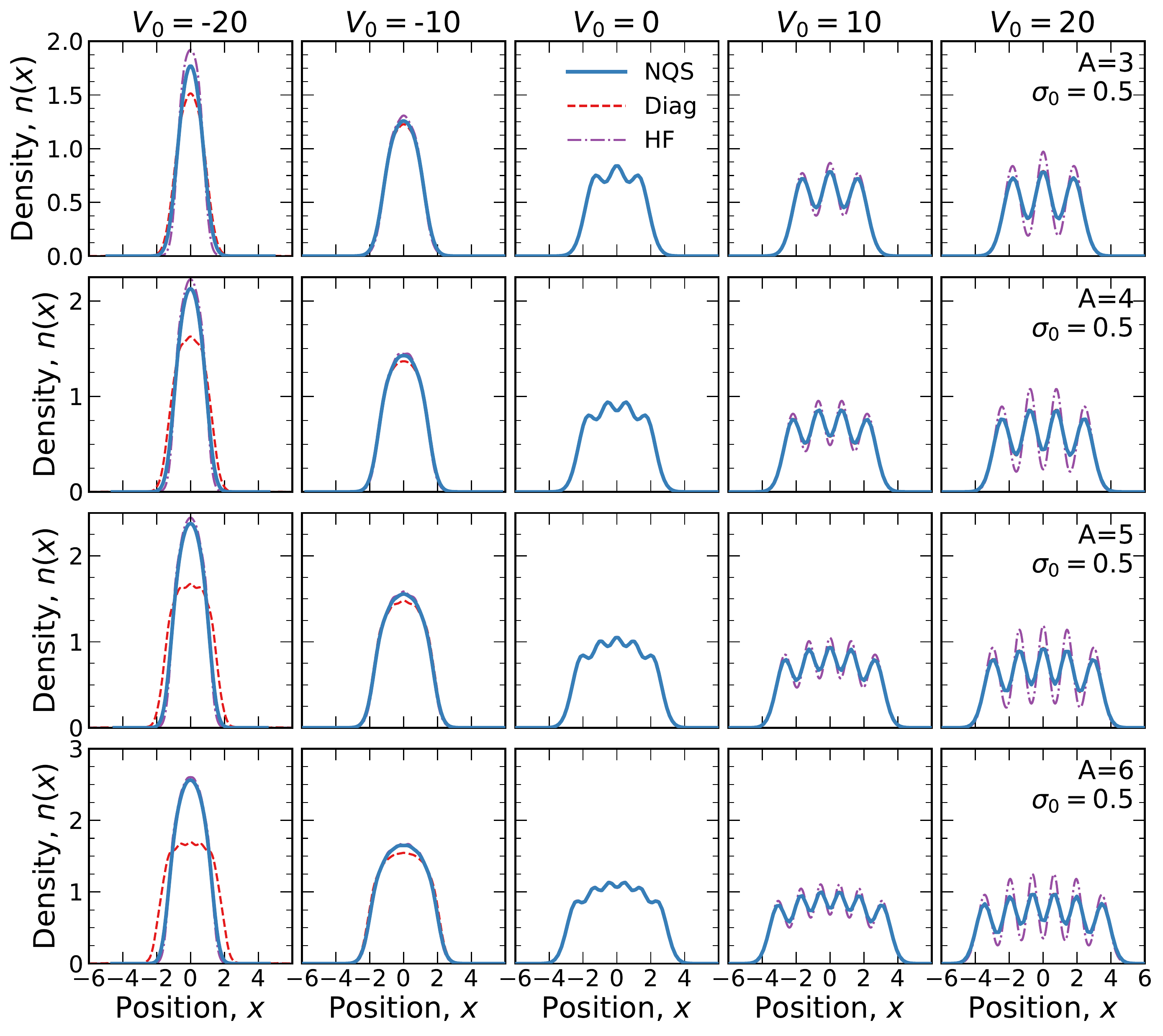}
\caption{Density profiles $n(x)$ as a function of position $x$ for different values of $V_0$ and a fixed
range $\sigma_0=0.5$. From left to right, $V_0$ goes from $-20$ to $20$ in
increments of $10$ units. From top to bottom, we show the results for $A=3$, $4$, $5$ and $6$ particles.}
\label{fig:density_profiles}
\end{figure*}

In terms of benchmarks, we notice two different remarkable tendencies. 
First, the HF results overestimate the ground-state energy, as expected due 
to the lack of backflow. The 
difference between the HF and the NQS prediction becomes less noticeable 
as the particle number increases. In other words, the HF approximation becomes
more reliable in relative energy terms as the number of particles increases.
This is not so surprising, since the mean-field picture should work better as the number
of particles increases.
Second, we find that the exact diagonalization technique is always less bound
than the NQS prediction for values $V_0<-10$. We interpret this as a 
consequence of the finite model space, which is insufficient in this strongly
attractive regime. 
Anticipating the results of Fig.~\ref{fig:density_profiles}, in the bosonized 
regime the density of the system is
substantially more compressed than in the noninteracting HO case. 
One may thus expect to
need many more single-particle modes to describe the  density profile and the energetics of the system.

To quantify further the importance of correlations in the system, we look at the 
correlation energy for the $A=3$ to $6$ particles as a function of $V_0$ 
in Fig.~\ref{fig:correlation_energies}.
We stress that the energy scale is the same for all panels. 
In other words, the difference between the HF and the full result appears to be relatively
constant independently of the particle number. On the repulsive regime at $V_0=20$, the energy increases by about $0.2$ units 
every time that we add one more particle to the system.  In contrast, the increase on 
the attractive size is less steep. 
We stress that in the NQS the nonzero correlation energy is a direct 
consequence of the backflow correlations in the \emph{Ansatz}. It is remarkable
that this extension allows the NQS to capture the asymmetric dependence
on correlation energy with $V_0$, due to the different phases of the system. 

In the attractive side, the direct diagonalization approach yields a relatively poor description
of the correlation energy. Nonphysical, positive values of $E_c$ are achieved for coupling constants below $V_0<-10$ or $-15$. On the repulsive size, we also find signs
of a relatively poorer description of the correlation energy with the direct diagonalization
approach for $A \approx 6$. Having said that, 
it is clear that
the direct diagonalization approach works much better in the repulsive side, where deviations from
the NQS prediction are within $0.05$ in absolute
energy terms. 

\subsubsection{Density profiles}

It is also very useful to inspect the density profiles, $n(x)$,
of the systems with $A=3$ to $6$ particles. 
We summarize these results in Fig.~\ref{fig:density_profiles},
where the columns correspond to
different interaction strengths and the rows correspond
to different numbers of particles.
We compare predictions from the NQS (solid), 
the direct diagonalisation approach (dashed)
and the HF approximation (short-dash-dotted lines). 

The middle panels, corresponding to the noninteracting case ($V_0=0$), show $A$ peaks on top of a Gaussian-like overall
behavior. This is an expected behavior, which arises as a result
of Eqs.~(\ref{eq:HOwfs}) and~(\ref{eq:OBDM_HF}). 
In the attractive regime, the peak-like
fermionic structure in the density profile disappears from
$V_0 \approx -10$ onwards. 
The NQS results show a single peak that becomes narrower
as $V_0$ becomes more attractive. This is in line with the HF
prediction, which approaches the NQS result as the number of
particles increases. In contrast, the direct diagonalization 
approach is unable to grasp the narrow density distribution
beyond $V_0 \leq -10$. 
At $V_0=-20$, the NQS and HF density  distributions are almost
a factor of $2$ narrower than the noninteracting case.
It is thus na\"ively expected that a model space truncation in the 
direct diagonalization cannot capture such strong 
density rearrangements. 

In the strongly repulsive limit, $V_0 \gg 0$, the interaction
effectively separates the fermions apart, leading to the full crystallization of
the system~\cite{Jauregui1993,DinhDuy2020}. 
The separation of single-particle orbitals, however, cannot proceed indefinitely because the harmonic trap is more and more effective as $|x|$ increases. 
While the position of the $A$ density peaks barely changes, the troughs between them become more and more defined as $V_0$ increases. 
In this repulsive regime, the direct diagonalization  for the density profile results coincide with the NQS predictions. In contrast, the HF approach fails substantially. The HF density profiles overestimate (underestimate) the peaks (troughs) and the disagreement becomes more pronounced as the number of particles increases. We stress that the differences in density profiles
arise from the existence of backflow correlations in our NQS \emph{Ansatz}. 
More details on the density profile and the average size of the system are provided in Ref.~\cite{Keeble_thesis}.

\begin{figure}
\includegraphics[width=0.8\linewidth]{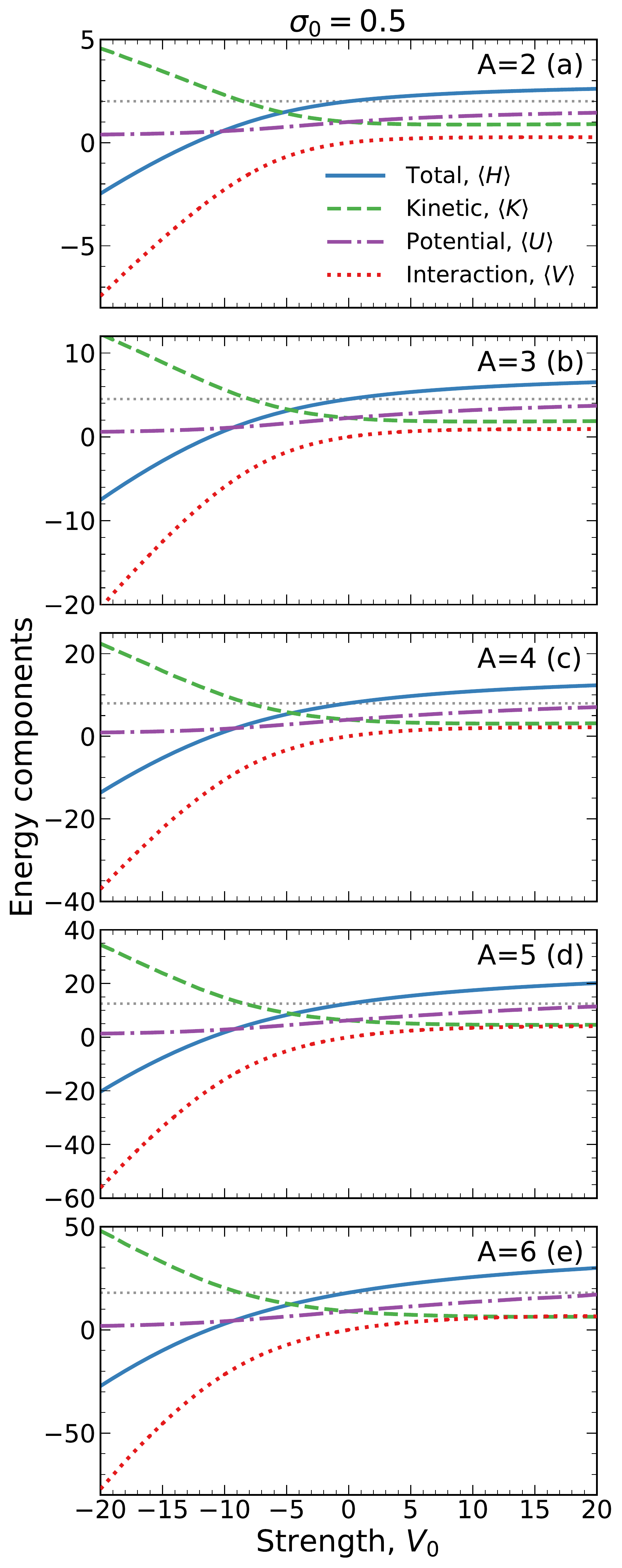}
\caption{Different components of the NQS ground-state energy  
as a function of interaction strength, $V_0$, for a fixed range $\sigma_0=0.5$. 
Solid, dashed, dash-dotted and dotted lines correspond to the total, kinetic,
harmonic oscillator and interaction energies, respectively. 
Panels (a) to (e) correspond to systems of $A=2$ to $A=6$ particles. 
The horizontal dotted line is the noninteracting ground-state energy.
}
\label{fig:energy_components}
\end{figure}

The difference of the density profiles in the attractive and repulsive 
regimes is reflected into the energetics of the system.  
We show different contributions to the total energy 
as a function of the interaction strength $V_0$
in Fig.~\ref{fig:energy_components}. Panels from top to bottom
correspond to different numbers of particles, from $A=2$ (top) to $A=6$ (bottom). 
We distinguish the contributions due to the 
kinetic energy, 
$\langle K \rangle = \langle \Psi \vert \sum_i - \frac{ \partial_i^2 }{2m} \vert \Psi \rangle / \langle \Psi \vert \Psi \rangle $;
the (external) harmonic oscillator potential energy,
$\langle U \rangle  = \langle \Psi \vert \sum_i - \frac{ x_i^2 }{2} \vert \Psi \rangle/\langle \Psi \vert \Psi \rangle$;
and the interaction energy, 
$\langle V \rangle = 
V_0/(\sqrt{2 \pi} \sigma_0)
\langle \Psi \vert \sum_{i<j} e^{- ( x_i^2-x_j^2 )/2 \sigma_0^2 } \vert \Psi \rangle / 
\langle \Psi \vert \Psi \rangle$. 
These components are all computed stochastically using the
NQS wave function probability. 

We find a common picture, which is relatively independent of
the particle number. In the noninteracting case,
the virial theorem stipulates that $\langle E \rangle = 2 \langle K \rangle=2 \langle U \rangle$ (and $\langle V \rangle=0$).
In the repulsive side, 
the harmonic oscillator and the interaction energy components increase 
very slowly as $V_0$ grows.  
We interpret this slow growth in terms of localization. When 
the single-particle orbitals become localized into
$A$ well-defined, equidistant peaks, the sharpening
of the fermionic features only modifies these two components
slightly. In contrast, the kinetic energy reduces substantially
as a function of $V_0$. In fact, for particle numbers
$A>4$, we observe that 
$\langle K \rangle \leq \langle V \rangle$, a condition that is 
typically employed to characterize a (Wigner) crystal~\cite{DinhDuy2020}.

The analysis in terms of different energy components indicates that 
the energy in the attractive regime behaves rather differently. 
First of all,
we find that the attractive total energy in the regime where 
$V_0 \ll 0$ is the result of the cancellation of two
large but opposite energy components.
On the one hand, the kinetic energy $\langle K \rangle$ increases substantially as 
$V_0$ decreases, as a result of the strong density rearrangement. 
On the other, the interaction energy $\langle V \rangle$ becomes extremely attractive
as all the particles are confined near the center of the trap and,
consequently, near each other. The central confinement also leads to
a near cancellation of the harmonic-oscillator potential energy $\langle U \rangle$. 

At this stage, 
we can draw some conclusions about the nature of the correlations in the repulsive
and attractive sides of the spectrum. Clearly, the bosonic transition in the attractive
side leads to a very strong rearrangement of the density profile. This rearrangement is so strong, in fact, that
the direct diagonalization based on noninteracting single-particle states struggles to capture it. 
An accurate description of this transition requires instead the solution of the system in real (or, possibly, momentum) space. We note that qualitatively similar results for the density profiles of $A=3$ and $4$ spinless fermions
are reported by Schilling et al.~\cite{Schilling2016} employing 
an even parity harmonium model. A universal bosonic peak for attractive interactions was observed in that case too.

In contrast, the rearrangement
of the density in the repulsive regime mostly leads to a localization of single-particle states that exaggerates the associated fermionic features. 
This is also found in the few-body system with an harmonium model~\cite{Schilling2016}.
The HF approach in this limit, however, overestimates the fermionic
features, which again highlights the importance of backflow correlations. 

\begin{figure}[t]
\includegraphics[width=\linewidth]{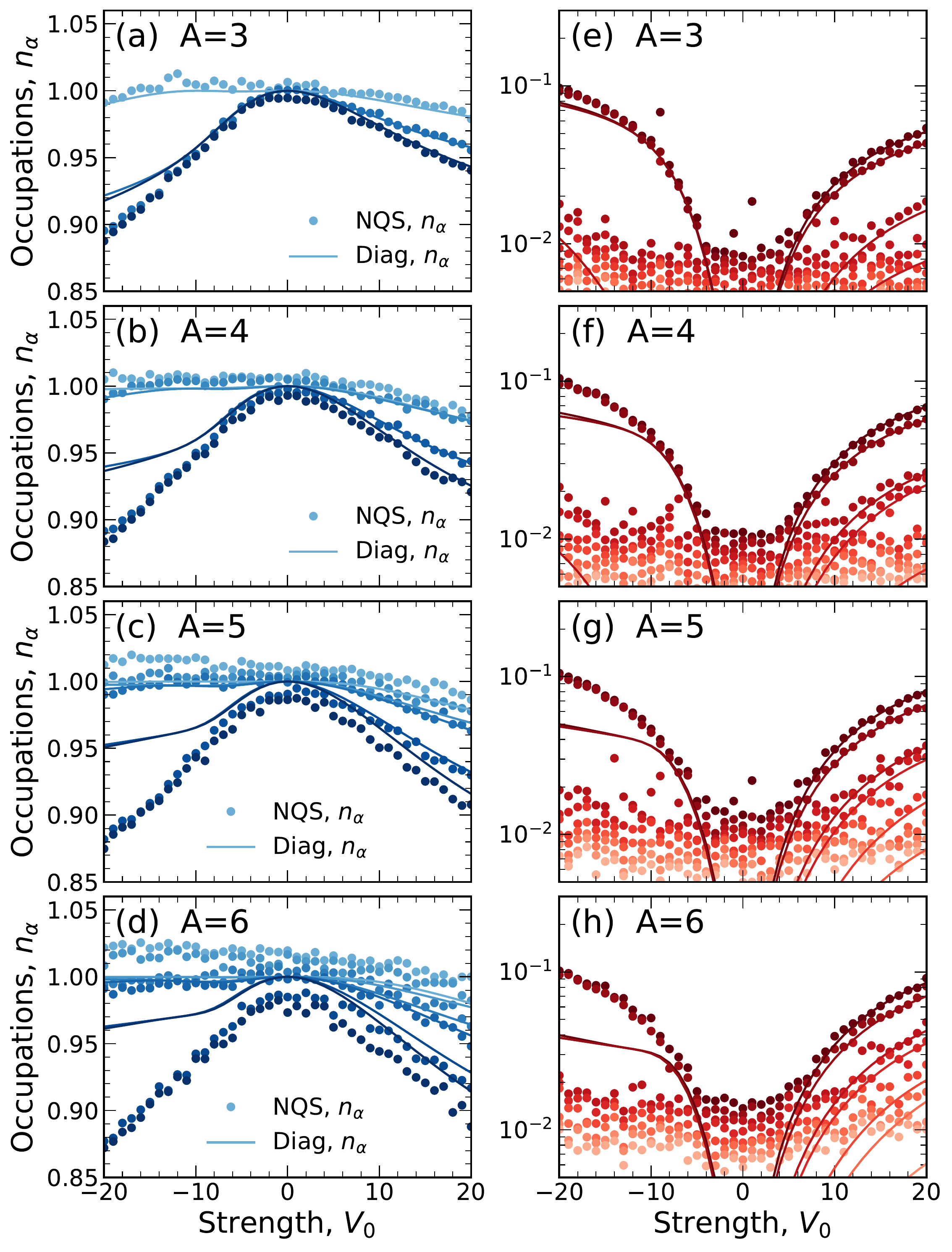}
\caption{
Panels (a)-(d): hole ($\alpha<A$) occupation probabilities for the $A=3$ (top row) to the $A=6$ (bottom row) systems as a function of interaction strength $V_0$ for a fixed range $\sigma_0=0.5$. 
We compare NQS results (solid circles) to the direct diagonalisation predictions (solid lines).
Panels (e)-(h): the same, in logarithmic scale, for particle ($\alpha \ge A$) states up to $\alpha=14$.
}
\label{fig:occupation-numbers_3plus}
\end{figure}

\subsubsection{Occupation numbers}

We get further insight on the correlation structure of the system by looking at the 
occupation numbers, $n_\alpha$, for different single-particle
states $\alpha$.  
We show $n_\alpha$ for systems with $A=3$ (top row) to 
$A=6$ (bottom row) in 
Fig.~\ref{fig:occupation-numbers_3plus}. The left (right) hand panels correspond to 
the occupation probabilities of holes (particles),  $n_{\alpha<A}$ ($n_{\alpha \geq A}$). 
There are distinct common structures appearing in the occupation numbers in the attractive
and the repulsive regimes. 
In the attractive case, the NQS predicts the appearance of one doubly degenerate state
that becomes substantially depleted as $V_0$ becomes more and more negative.  
This result is commensurate with the direct diagonalization occupation prediction, which also appears
to be double degenerate, although, somewhat less depleted. We interpret the difference between the NQS and the direct diagonalization prediction,
again, as a sign of the truncated model space. 
Both the NQS and the diagonalization predictions indicate that the remaining $A-2$ hole states are fully occupied to within a $2 \%$ accuracy, although, the statistical 
noise floor in the NQS approach makes it difficult to quantify this statement. 

The particle states shown in Figs.~\ref{fig:occupation-numbers_3plus}(e)-\ref{fig:occupation-numbers_3plus}(h) show the appearance of a doubly-degenerate 
state in the repulsive regime for all values of $A$. According to the NQS
prediction, these states reach an occupation of $\approx 10 \%$ at $V_0=-20$.
These doubly-degenerate states are reminiscent of a bosonization of the fermionic
degrees of freedom. This happens in spite of the absence of active spins
and hence can only be observed in the case of a finite-range interaction. Doubly degenerate structures in the attractive regime have also been observed in variational calculation of one-dimensional (1D) $p$-wave fermions~\cite{Koksik2020}. These simulations, however, have a harder time describing the repulsive regime, which the NQS can access seamlessly. 

In the repulsive regime, $V_0 > 0$, no double degeneracy is observed. Single-hole states [Figs.~\ref{fig:occupation-numbers_3plus}(a)-\ref{fig:occupation-numbers_3plus}(d)] 
and single-particle states [Figs.~\ref{fig:occupation-numbers_3plus}(e)-\ref{fig:occupation-numbers_3plus}(h)] are all distinct, and one
can clearly distinguish $A$ states on the repulsive branch of each panel
with occupations $n_\alpha > 5 \times 10^{-3}$. This
prediction is supported both by the direct diagonalization and the NQS results.
For the $A=3$ system, Fig.~\ref{fig:occupation-numbers_3plus}(a), we find that three distinct single-hole states appear 
with small depletions, with occupations of more than $95\%$ up to $V_0 \approx 20$. 
The corresponding single-particles states in Fig.~\ref{fig:occupation-numbers_3plus}(e) are clustered into two 
nearly degenerate states with populations of $\approx 5 \%$ at $V_0 \approx 20$,
and another state with a population of $2 \%$. A similar picture emerges as $A$ increases.
For even values of $A$, we find that there is a near (but not complete) degeneracy of 
$A/2$ states. In the single-hole cases, the occupations decrease steadily with $V_0$
in pairs. Similarly, the occupation of particle states increases in pairs of similar
values. For odd $A$, the picture is essentially the same, except that there is an odd
single-particle orbital with a specific occupation probability. We note that there
are similarities between this picture and the results obtained in Ref.~\cite{Schilling2016}
for 1D spin-polarized trapped fermions with harmonic pair interactions.

For both the single-particle and single-hole states, the statistical noise floor is on the order of $\approx 10^{-2}$. We find a slight increase of this floor with particle number, of
about a factor of $2$ when moving from $A=3$ to $A=6$. 
Although one can see the statistical noise within the single-particle states, 
their effect is even more evident in the
single-hole states, where they lead to nonphysical fluctuations with $n_\alpha>1$. 
We arbitrarily limit the number of single-particle states to $14$ within this analysis.
The noise floor is consistent for all values of $V_0$, indicating that 
its source stems from the stochastic methodology used to calculate the OBDM. 
We limit the range of occupation values shown for the single-hole states to be above $5 \times 10^{-3}$, which highlights a few of the lowest single-hole states obtained with the direct diagonalization approach (solid lines). 

\begin{figure*}[t] 
\includegraphics[width=0.6\linewidth]{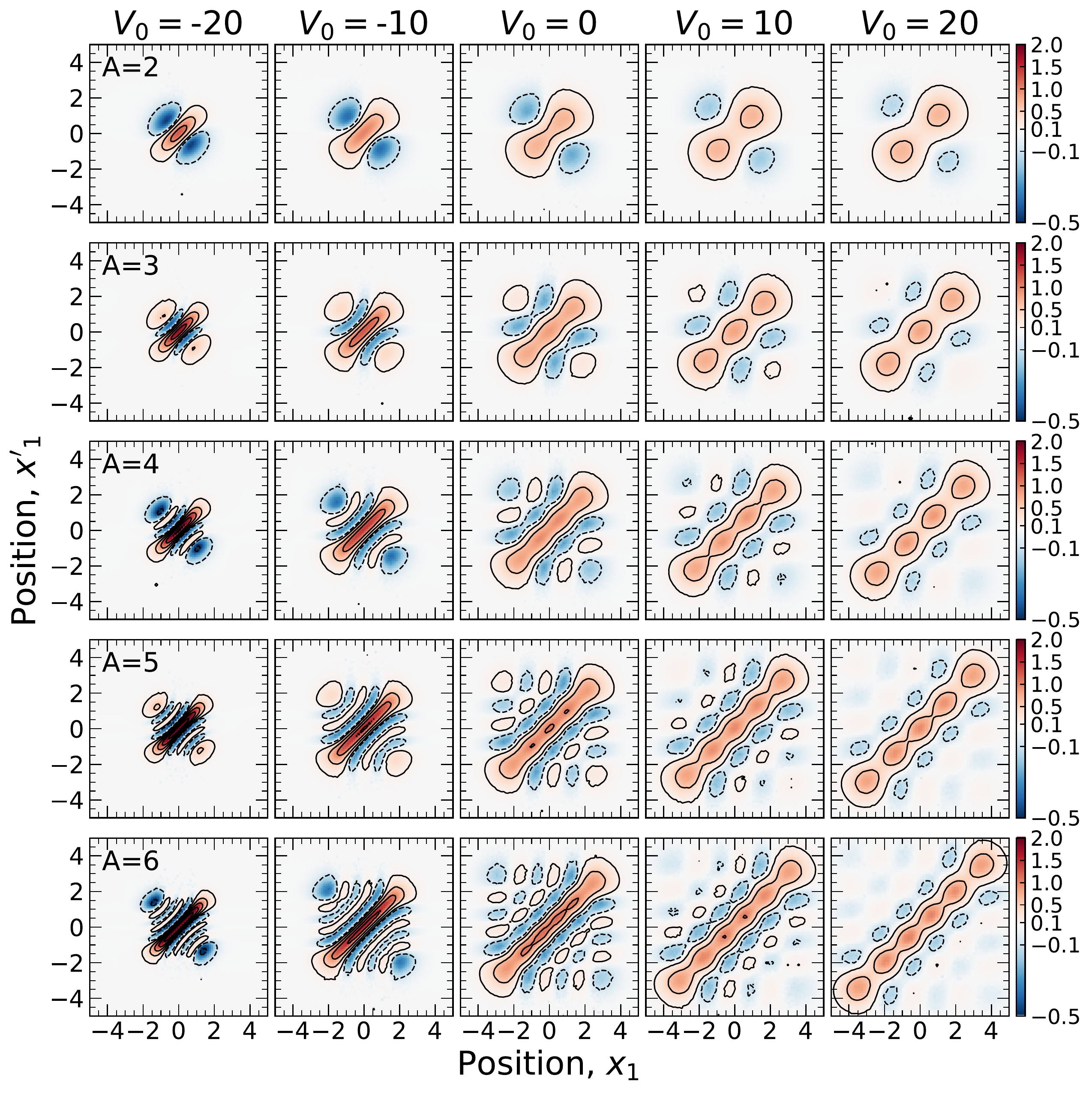}
\caption{Density contour plots of the one-body density matrix $\rho(x_1',x_1)$ of systems from $A=2$ (top row) to $A=6$ (bottom row) particles. 
Columns correspond to different interaction strengths, from $V_0=-20$
(left column), to $V_0=20$ (right).
Contours are shown at density matrix values of 
$-0.5$, $-0.1$, $0.1$, $0.5$, $1.0$, $1.5$ and $2.0$. The colour coding is the same across all panels.
}
\label{fig:denmat}
\end{figure*}

\subsubsection{One-body density matrix}

So far, we have discussed local properties of the system. We now turn 
our attention to nonlocal structures reflected in the density
matrices. 
We discuss the NQS results in this and the following sections,
and refer the reader to the Appendix for the analogous 
HF and direct diagonalization results.
In Fig.~\ref{fig:denmat}, we show the OBDM $\rho(x_1',x_1)$ 
of the ground-state wave function for $A=2$ (top row) to 
$A=6$ (bottom row). 
The different columns correspond to interaction strengths 
values from $V_0=-20$ to $20$ (in steps of $10$). 
The central panels correspond to the noninteracting case, 
$V_0$, which can be understood in relatively simple analytical 
terms from a combination of Eq.~(\ref{eq:OBDM_HF}), valid
in the noninteracting case, and the HO eigenstates of 
Eq.~(\ref{eq:HOwfs}), leading to Eq.~(\ref{eq:OBDM_HO})~\cite{Rius2023}. 
In all cases, the density matrix 
is purely real, and covers a more-or-less square area 
in the $x_1'-x_1$ plane. 
This is a consequence of the combination of
single-particle terms $\varphi_n^*(x_1') \varphi_n(x_1)$ in the sum of
Eq.~(\ref{eq:OBDM_HF}), which naturally provide a limit both within
and outside the diagonal of the $x_1'-x_1$ plane~\cite{Rios2011}.
This is clearly seen in Eq.~(\ref{eq:OBDM_HO}), which indicates that 
$\rho(x_1',x_1)$ is given by a Gaussian
envelope in the $x_1'-x_1$ modulated by a polynomial. 

Moreover, one finds that there is a relatively narrow area of 
positive values near the diagonal, where $\rho(x'=x,x) \equiv n(x)>0,$
is positive definite. As one moves away from the diagonal,
definite ripples appear in areas with consecutive positive and
negative values. These ripples reflect the nodal structure of the
single-particle states. In fact, there are as many
changes of sign along the $x_1'=-x_1$ direction in one side of the diagonal 
as particles in the system.
Equation (\ref{eq:OBDM_HO}) provides an explanation of this in terms of the 
roots of the polynomials of order $2(A-1)$ along the $x_1'=-x_1$ direction. 

In the attractive case, shown in the left panels, the overall size
extent of the system decreases (see Fig.~\ref{fig:density_profiles}). 
Similarly, the support of the OBDM in the $x_1'-x_1$ plane also 
diminishes. Moreover, the peaks and troughs associated to the changes
of sign in the off-diagonal direction move closer to the diagonal. 
Importantly, these interference effects appear to increase in magnitude
as the interaction strength becomes more negative. We stress that the
diagonalization of this strongly oscillating OBDM leads to occupation
numbers with a dominant doubly degenerate structure, more reminiscent of
a coherent picture. 

In the strongly repulsive case, in the panels on the right, 
one can clearly distinguish the localization of single-particle
orbitals along the diagonal in the form of well-defined 
equidistant peaks. The off-diagonal oscillations and change of
signs observed in the non-interactive and the attractive regimes,
are substantially damped in this case. At $V_0=20$, for instance,
one only observes a well-defined series of $A-1$ quasicircular regions 
with negative values in a direction parallel to the diagonal. 
While there are additional oscillations along the off-diagonal
direction, these have a much smaller amplitude, to the point that
they can hardly be distinguished in the scale of the figure. 

\begin{figure*}[t] 
\includegraphics[width=0.6\linewidth]{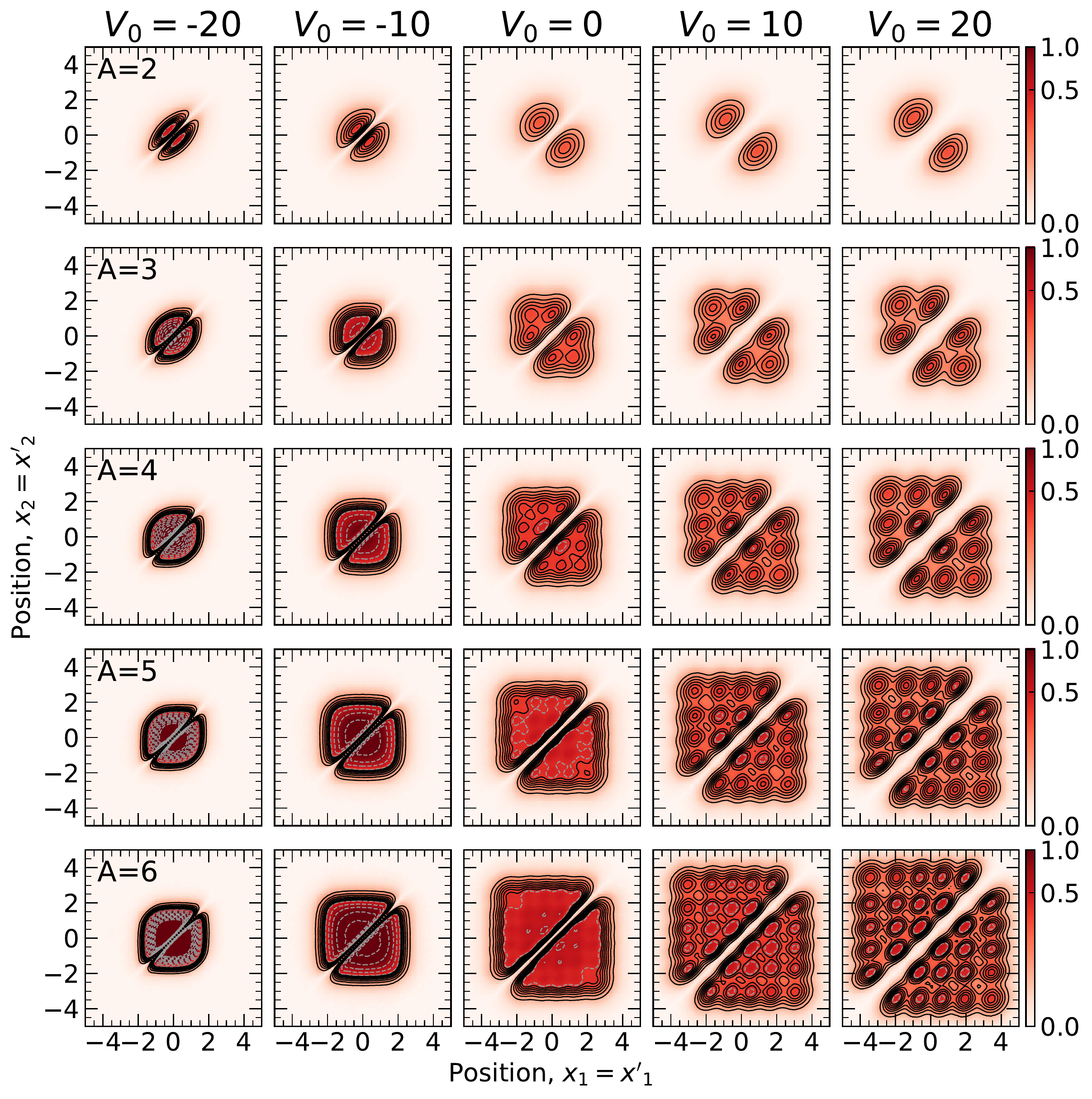}
\caption{Density contour plots of the pair correlation function $g(x_1,x_2)$ for systems from 
$A=2$ (top row) to $A=6$ (bottom row) particles. 
Columns correspond to different interaction strengths, from $V_0=-20$
(left column), to $V_0=20$ (right).
Contours are shown at steps of $0.05$ from $0$ to $0.35$ (solid lines), 
and at steps of $0.2$ from $0.4$ to $2$ (dashed lines). 
The colour coding is the same across all panels.
\label{fig:TBDM}
}
\end{figure*}

\subsubsection{Pair correlation function}

Finally, we present in Fig.~\ref{fig:TBDM}, the PCF, $g(x_1,x_2)$
of Eq.~(\ref{eq:g_pair_dist}) for different particle numbers. 
The panels are organized
in the same way as Fig.~\ref{fig:denmat}. 
The PCF provides information of correlations due both to the 
antisymmetric nature of the system and to the presence of interactions~\cite{Knight2022}. 
The function $g(x_1,x_2)$ is very different in the attractive and the repulsive
regimes, but there are common structures across all panels. 
Specifically, we find a ridge
around $x_1 \approx x_2$ where the pair correlations vanishes.
This is a consequence of the Pauli exclusion principle, which in the spinless
fermion case imposes the condition $g(x,x)=0$. 
In the noninteracting system, the condition is enforced by the $(x_1-x_2)^2$ term of
Eq.~(\ref{eq:correlation_function_HO}). 
The width of this ``Pauli ridge'', however, changes substantially 
with interaction strength, which indicates that part of it is also a consequence
of interaction effects. On the repulsive
side, shown on the panels to the right of the center, one clearly finds that the
ridge becomes wider with $V_0$. This effect is a reflection of the spatial localization
of the orbitals observed in Figs.~\ref{fig:density_A2} and \ref{fig:density_profiles}. 
We expect the size of the ridge in the repulsive side to be sensitive to both
$V_0$ and the range of the interaction, $\sigma_0$. 
As $V_0$ is more repulsive, particles try to localize more, leading to more distant and well-defined peaks in the PCF.

In contrast, on the attractive regime displayed
on the left panels of the figure, 
we find that the ridge becomes narrower as $V_0$ becomes more attractive, to 
the point that, for $A>5$, the ridge is difficult to see on the leftmost panels
with $V_0=-20$. The pair distribution function in the attractive side is relatively 
featureless and, other than the ridge, it is extremely peaked in an almost square 
region around the center. This indicates a high likelihood of finding one
particle in a region within a few units of position of the center, a picture
that is reminiscent of a Bose-Einstein condensate. This picture is in line with the PCF of two spinless fermions with a soft-core potential~\cite{Koksik2018} and of the correlation structure of P-wave fermions observed in Ref.~\cite{Koksik2020}.

In contrast with this relatively featureless peak, $g(x_1,x_2)$ on the repulsive side has a much richer
structure, with several well-defined peaks. These peaks are clear indications
of the emerging localized structure of the system, and have been observed in analogous
simulations of one-dimensional, few-body Wigner crystals~\cite{Jauregui1993}. For odd
particle numbers, we find that there is always a peak to the right (or left) of the
Pauli ridge along the $x_1=0$ direction. For these systems, the density has a peak at
$x=0$ (see Fig.~\ref{fig:density_profiles}), 
so one particle sitting at $x_1=0$ will have a large probability of 
finding $A-1$
particles at fixed intervals in $x_2$. For $A=3$, this occurs at $x_2=\pm 2$, whereas
for $A=5$ this happens at about $x_2=\pm 1.5$ and $\pm 3$.

The structure for an even number of particles is relatively different.
For an even number of particles, there is a depletion in the density distribution at the center
of the trap, see Figs.~\ref{fig:density_A2} and \ref{fig:density_profiles}. Consequently, the
PCF along $x_i=0$ does not show any peaks. Instead, the lines of horizontal or 
vertical $A-1$ peaks 
are found at around $x_i \approx \pm 1$. 
Whereas for $A=4$ particles a second set of three peaks is found at 
$x_i = \pm 2.5$, for $A=6$  particles this is localized at $x_i=\pm 2$.
The well-defined single-particle peaks are therefore more compressed as 
$A$ increases, in line with the results of Fig.~\ref{fig:density_profiles}.

All in all, the PCF for all 
systems indicates a well-defined localized phase in the repulsive regime, and a Bose-like phase in 
the attractive regime. This complex behavior is obtained with an NQS formed by a single determinant 
with backflow correlations. In other words, for a fixed particle number, the same NQS is capable of
describing systems with very different spatial extents, correlation structures and underlying physical pictures. We take this as a promising sign for
the performance of NQS approaches in the description of condensed matter
systems.

\section{Conclusion and outlook}
\label{sec:conclusion}

To conclude, we have applied the NQS method to solve the Schr\"odinger equation for one-dimensional systems of $A$ harmonically trapped, interacting fermions without spin degrees of freedom. We model the interaction using a toy model Gaussian form factor, motivated by nuclear physics applications and such that, in the zero-range limit, we recover the noninteracting case. 
We purposely choose a set of interaction strengths and ranges where perturbation theory may not converge easily. In particular, we find that an interesting choice for the interaction range is half the natural oscillator length. 

The NQS that we employ provides a fully antisymmetric \emph{Ansatz} to represent one-dimensional ``spinless'' fermions. The NQS has two equivariant layers of $H = 64$ hidden nodes each. These layers embed the one-dimensional positions of $A$ particles into a $H$-dimensional space. These layers are then projected into a single GSD. 
This approach benefits from backflow correlations from the outset, and does not explicitly require a Jastrow factor. The number of parameters of the NQS remains nearly constant when going from $A=2$ to $A=6$ particles. 
By combining the NQS with VMC techniques, we obtain a wave function that is able to represent many-body correlations and can accurately describe very different physical regimes without any specific modifications to the network architecture. 

The focus of our paper is on benchmarking the new NQS approach with other well-known many-body methods. For all values of $A$, we compare against the mean-field Hartree-Fock method and the direct diagonalization approach. Moreover, in the $A=2$ case we have access to an exact solution in real space as well as a perturbative treatment of the Gaussian interaction. The results for all values of $A$ indicate that the NQS can efficiently provide an accurate representation of the many-body wave function.
One benefit of the NQS approach formulated in the real-space continuum 
is its variational performance, that surpasses mean-field approaches and does not suffer from basis truncation effects, like the direct diagonalization method does. 
We characterize our benchmarking process using several properties, including the ground-state energy, density profiles, single-particle occupation numbers, as well as the one-body density matrix and the PCF. 

The NQS can accurately describe the ground-state wave function across different physical regimes. In particular, we find two very different pictures that arise as the strength of the interaction goes from very attractive to very repulsive values. 
We find a common, method-independent interpretation for these two phases 
of 1D trapped spinless fermionic systems. 
On the attractive side, the density 
distribution of the system is a single peak centered at the origin, reminiscent of a 
Bose-Einstein condensate. The peak becomes narrower and higher as the strength becomes
more negative. This represents a substantial density rearrangement in the system, that
is hard to capture with a truncated basis direct diagonalization approach based on an harmonic-oscillator basis. A real-space implementation of the Hartree-Fock method, however, describes well this phase for sufficiently large $A$.
In this regime, the one-body density matrix is strongly oscillating off
the diagonal and the pair distribution function is narrow and squared shaped. We find that a degenerate set of $2$ single-particle orbitals are 
strongly depleted, whereas the remaining $A-2$ states remain fully occupied.

\begin{figure*}[t] 
\includegraphics[width=0.6\linewidth]{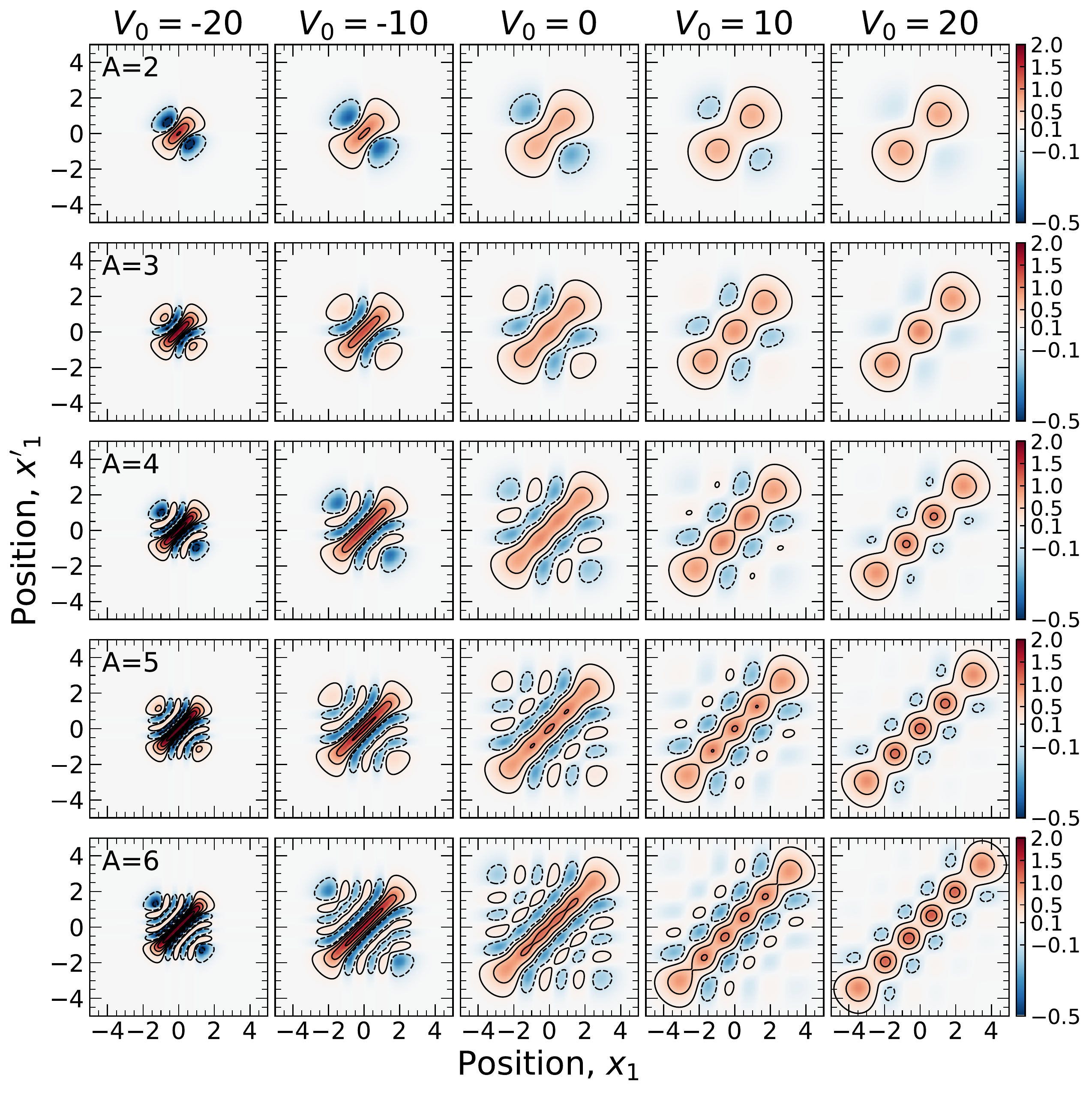}
\caption{The same as Fig.~\ref{fig:denmat} for the Hartree-Fock approximation.}
\label{fig:denmat_HF}
\end{figure*}

In contrast, for large and repulsive values of the interaction strength we find a completely different picture. In this
limit, we reach the regime of a localized crystal, and fermions spontaneously
order in a periodic fashion. This is reflected in the density distribution, 
which has a well-defined periodic structure, with $A$ peaks, as well as
the PCF, which shows associated crystalline
structures. In terms of occupation numbers, the repulsive regime exhibits
quasidegeneracies in pairs of orbitals. We note that this regime is well described
by the direct diagonalization approach, but the HF approximation predicts density 
distributions with exaggerated localization features. 

We foresee several applications of this work in the near future, and distinguish
between applied and formal developments. On the applied side, we have found clear
signatures of two physical phases in relatively simple fermionic systems. When 
spin is not present, finite-range interaction effects are necessary and lead
to very different behavior in the attractive and repulsive regime. 
Here, one may exploit NQSs to investigate physical features
that have not been explored yet, like higher-order density matrices.
Moreover, one could also employ the NQS wave function to compute
matrix elements of other relevant operators employing standard
Monte Carlo techniques~\cite{becca_sorella_2017}.
Having access to the full wave function of the system, we should also be 
able to characterize the evolution of entanglement measures as a function of 
interaction parameters. 
It would be interesting to perform NQS simulations with
odd-parity potentials, which resemble the more realistic 
interactions employed in the past in spinless fermion studies~\cite{Girardeau2003,Sowinski2019,Valiente2020}. 
We stress that there is no 
fundamental limitation hampering the applicability of the NQS method for
derivative odd-parity potentials.  
In addition, one could extend this approach to other trapped systems, with 
hard boundary conditions~\cite{Radu2022}, or to 
infinite systems incorporating periodic boundary conditions
\cite{Wilson2022,Pescia2023,Lim2023}.

On the formal side, there are interesting step forwards to take in a variety of directions. 
First, it would be interesting to quantify the quality of the wave function \emph{Ansatz} in more formal terms. In other words, one could try and gauge what information is lost when introducing this NQS \emph{Ansatz}.
A potential avenue may involve the quantification of entanglement measures
on NQS. 
In applications
of one-dimensional ultracold atoms, spin degrees of freedom are relevant, 
and often manipulated experimentally. 
This calls for an explicit, general spin-dependent representation of NQSs, 
along the lines
recently explored in lattice systems~\cite{RobledoMoreno2022},
nucleons~\cite{Lovato2022}, the electron gas~\cite{Wilson2022,Pescia2023} or unitary fermions~\cite{Lim2023}, but potentially applicable to SU($N$) 
interactions. A comparison with the rich variety of experiments in
nonperturbative regimes here may provide a fertile ground for tests
of the applicability of NQS \emph{Ans\"atze}. 
Finally, our findings indicate that an NQS originally 
pretrained in a noninteracting case can eventually learn at least two different phases of a quantum system. It would be interesting to analyze whether this result is general, and whether NQSs can indeed perform across complex quantum phase-transition boundaries.

Our ultimate aim is the study of nuclear physics systems employing NQSs. 
To this end, we need to extend our current technology to treat three-dimensional
systems. This also requires the inclusion of spin and isospin degrees of freedom. 
All studies so far indicate that nuclei in the few-body sector can be
theoretically accessed by employing 
NQSs~\cite{Keeble2020,Adams2021,Gnech2022,Lovato2022}. It remains to be seen
whether the promising scaling with particle number that we have found in this work can be efficiently transferred into a nuclear physics setting and deliver 
a substantial change in \emph{ab initio} nuclear theory.

\begin{figure*}[t] 
\includegraphics[width=0.7\linewidth]{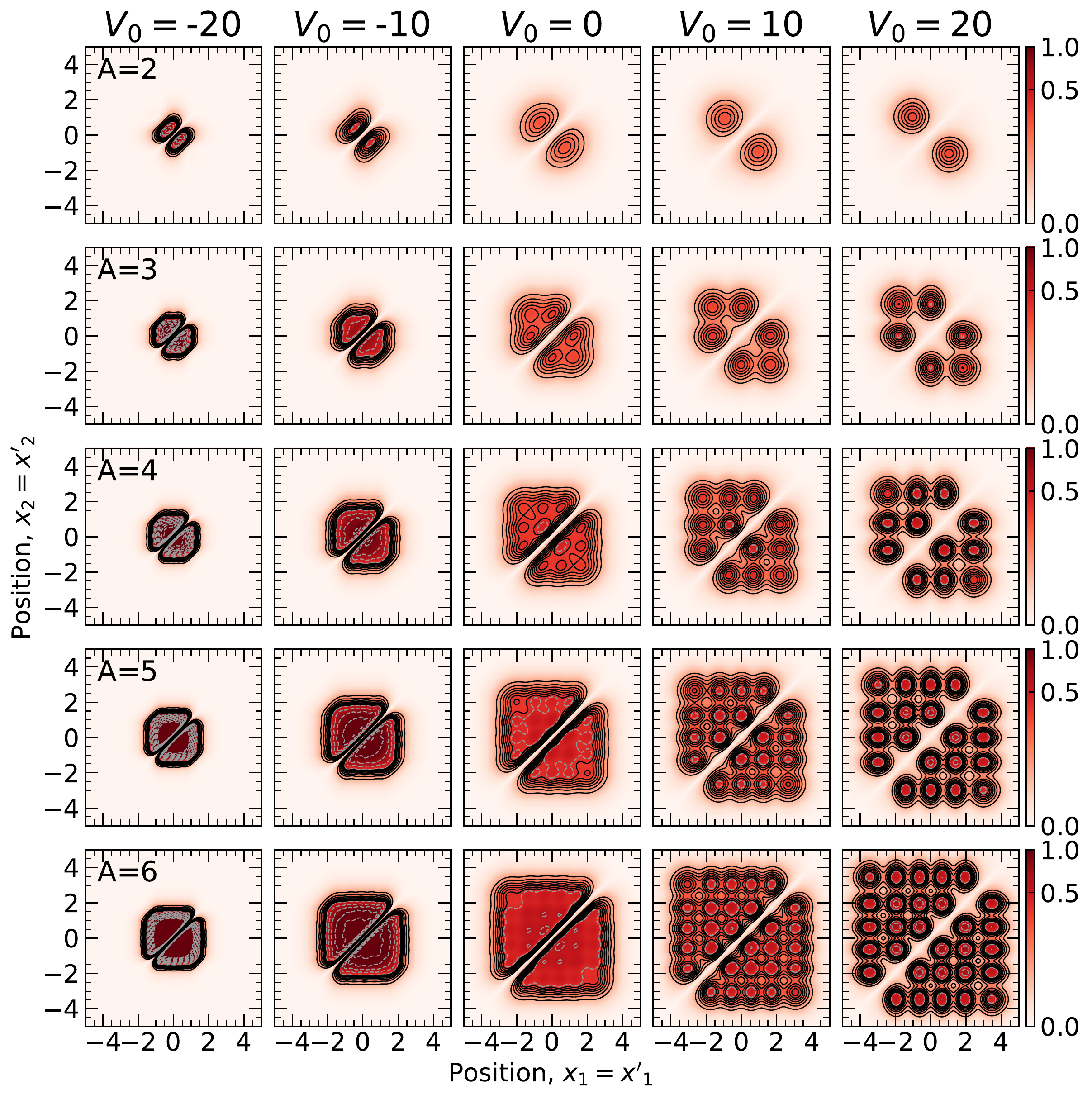}
\caption{The same as Fig.~\ref{fig:TBDM} for the Hartree-Fock approximation.}
\label{fig:TBDM_HF}
\end{figure*}

\appendix
\section{Non-local properties}

In this Appendix, we look at the nonlocal properties obtained with our benchmarks models. We provide, specifically, an analysis of the OBDM and PCF obtained within the HF approximation and within the direct diagonalization framework. On the one hand, this provides a useful cross-check with the NQS results of the main text. On the other, it allows us to identify the effect of correlations in the nonlocal properties of the system. 

\subsection{Hartree-Fock}

We start by discussing the results of Fig.~\ref{fig:denmat_HF}, which shows the
OBDM in the HF approximation. 
The format and panel structure is the same as shown for the NQS data
in Fig.~\ref{fig:denmat}: different rows correspond to particle numbers from
$A=2$ to $A=6$ and different columns indicate different interaction strengths for 
$\sigma_0=0.5$. 
Qualitatively, Figs.~\ref{fig:denmat} and \ref{fig:denmat_HF} are very similar. 
In
the attractive regime (panels to the left of the center), 
we find that the system becomes more compact as
$V_0$ becomes more negative. Just as in the NQS case, the HF 
solution shows a strong oscillatory behavior off the diagonal. In contrast to
the NQS solution, however, the central ridge in the HF results is higher 
and the overall support of the density matrix has a more well-defined, square shape.

The central column shows the same results for the NQS and the HF solution, as expected 
since this is a noninteracting case. 
Nonetheless, this is a nontrivial check on the numerical accuracy of the NQS solution, which
thus reproduces not only the energy but also the nonlocal properties of the noninteracting
wave function. 
On the repulsive side, shown in the panels to the right of the center,
the HF results are also very similar to the NQS data shown in 
Fig.~\ref{fig:denmat}. In the HF solution, we find that there is more structure 
concentrated along the 
diagonal, with more localized solutions and clearly identifiable peaks compared to the
NQS simulations. This bodes well with the findings of the density profiles 
shown in Fig.~\ref{fig:density_profiles}.
Off the diagonal, the NQS results show a slightly stronger oscillatory behavior
than the HF predictions. 

We now discuss the PCF within the HF 
approximation, which we show in Fig.~\ref{fig:TBDM_HF}. 
Here, the format is the same as Fig.~\ref{fig:TBDM} in the main body of the text.
Just as in the case of the OBDM, the central column of the two cases is the same.
This indicates, again, that the noninteracting case is under control with the NQS \emph{Ansatz}. This benchmark indicates that the NQS \emph{Ansatz} does not lose any significant amount of information at the two-body level.

Differences between the HF and NQS pair distribution
functions can be observed in both the attractive and repulsive regimes. In the attractive case, the support for the HF correlation functions is smaller than in the NQS case. This is in line with the corresponding density
distributions, which are narrower in the HF case. In
addition, the HF case shows a sharper, square-like behavior in the attractive cases and a somewhat narrower Pauli ridge. 

\begin{figure*}[t] 
\includegraphics[width=0.6\linewidth]{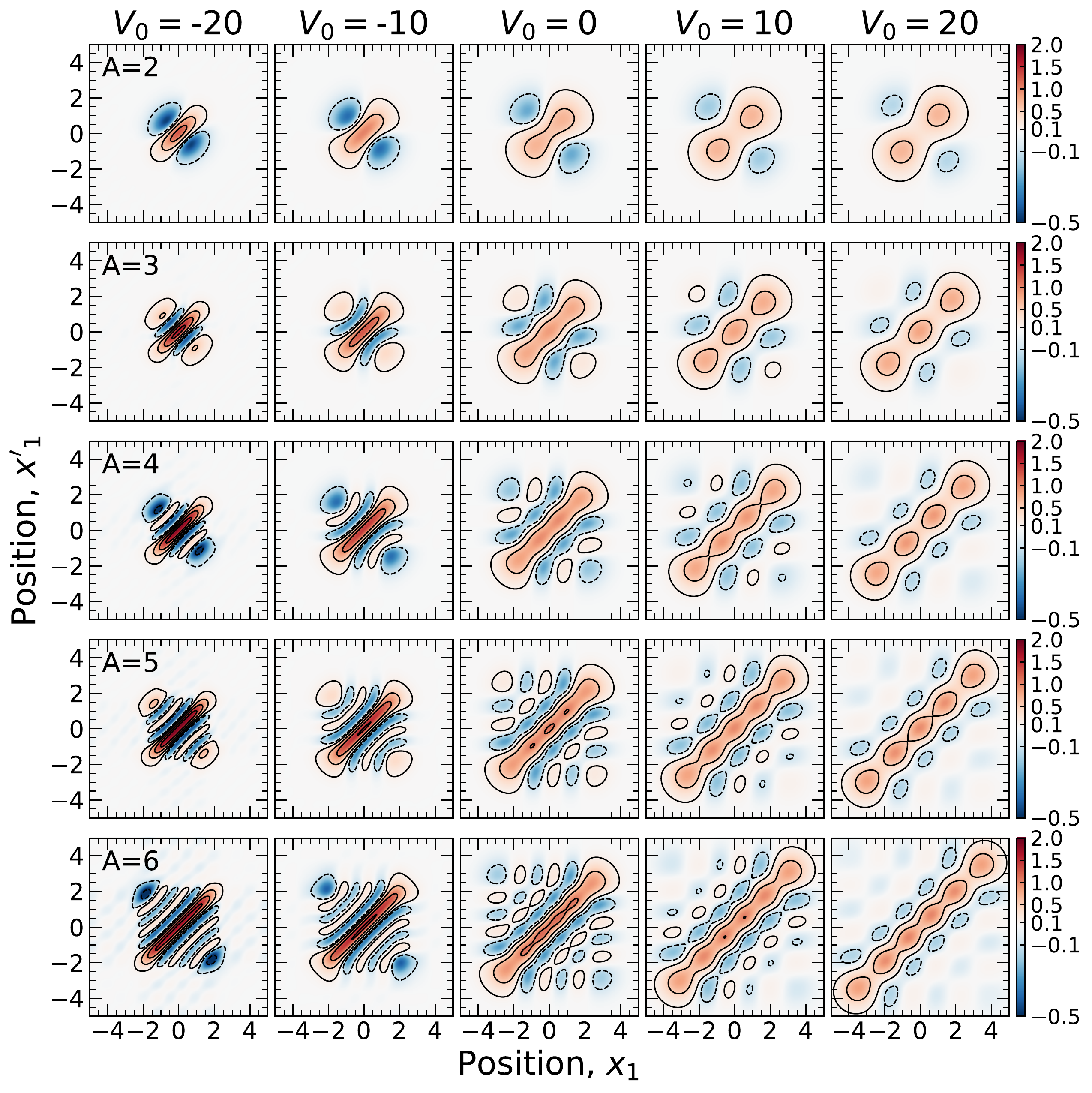}
\caption{The same as Fig.~\ref{fig:denmat} for the direct diagonalization approach.}
\label{fig:denmat_diag}
\end{figure*}

In the repulsive case, the HF densities show a much more 
localized state. In the pair distribution function, this
translates into very well-defined peaks. The peaks in the HF distribution lie close to those predicted by the NQS,
but are generally rounder and higher.

\subsection{Direct diagonalization}

We have also computed the OBDM and the PCF for the direct diagonalization simulations. We already know from the analysis of the energetics, the 
local densities and the occupation numbers that the exact
diagonalization solution has convergence issues in the attractive regime. We therefore expect to see differences 
in this sector, which should be taken as  
indications of the lack of numerical convergence associated with the
truncation of the finite model space 
as opposed to any actual physical limitations.

We show in Fig.~\ref{fig:denmat_diag} the OBDM obtained from the exact diagonalization approaches in the same format as Figs.~\ref{fig:denmat} and ~\ref{fig:denmat_HF}. As expected, there are virtually no observable differences in the central and right panels of the Figures. In this regime, the exact diagonalization results are well converged and the agreement with the NQS can be taken as a solid qualitative benchmark. In other words, the one-body 
nonlocal properties of the NQS wave function are well under control. 

On the attractive regime, in the panels left of the center, we observe increasing differences as the interaction strength $V_0$ and the particle number $A$ increase. The lowest-left panel, with $V_0=-20$ and $A=6$, is the most affected in this case. We find a density matrix with extended support in the $x_1-x_1'$ plane. Off the diagonal, we observe the same number of oscillations in the NQS and the exact diagonalization cases, but the latter shows a larger, less compact oscillation structure. We assume that this is due to the absence of high energy modes in the active space, which may be able to account for faster off-diagonal oscillations.  

\begin{figure*}[t] 
\includegraphics[width=0.6\linewidth]{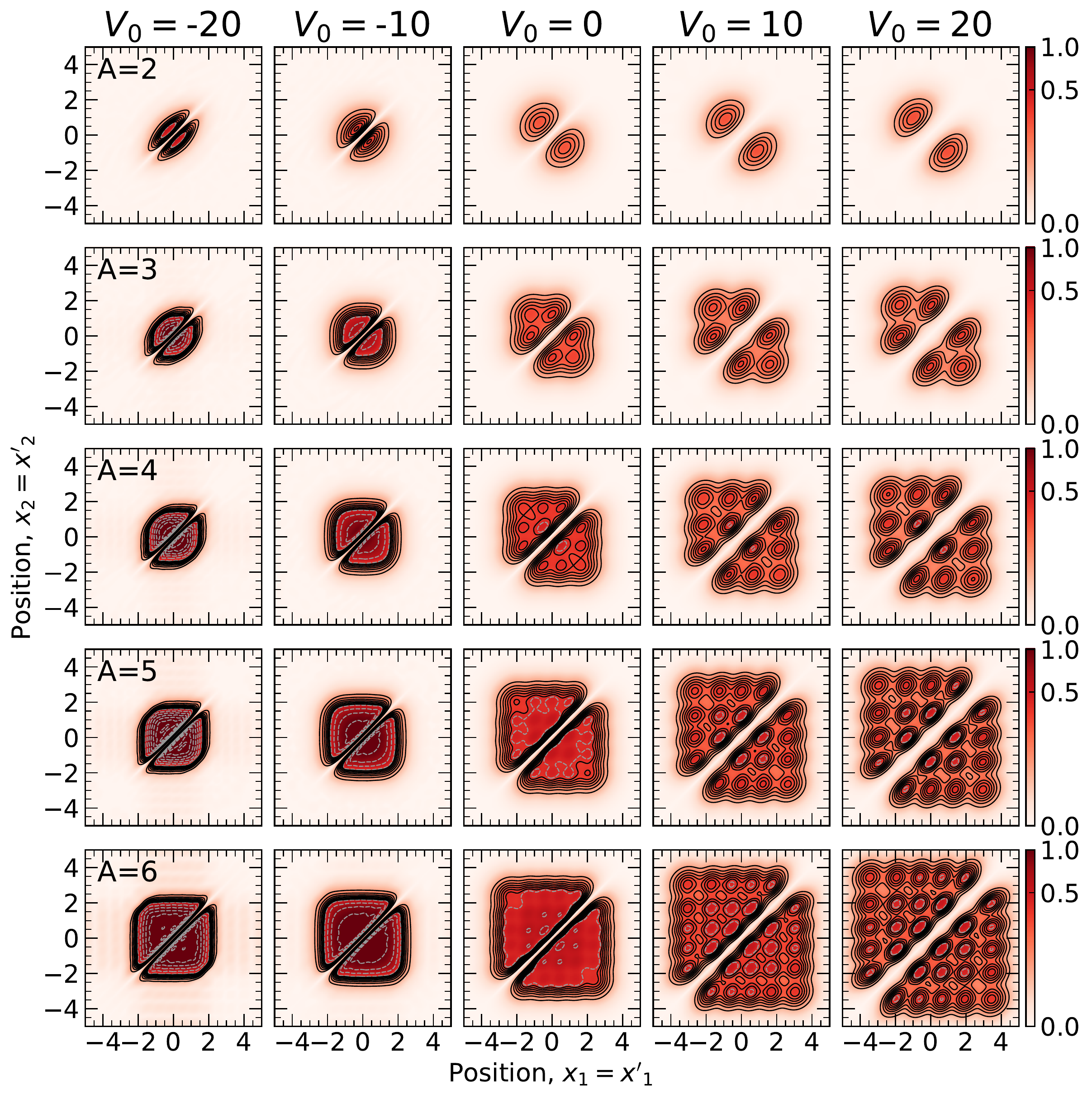}
\caption{The same as Fig.~\ref{fig:TBDM} for the direct diagonalization approach.}
\label{fig:TBDM_diag}
\end{figure*}

Finally, Fig.~\ref{fig:TBDM_diag} shows the PCF extracted from the exact diagonalization approach. On the noninteracting and the repulsive columns, there are no notable differences between the NQS results of  Fig.~\ref{fig:TBDM} and the direct diagonalization results. Again, this can be taken as an indication of the high-quality NQS wave function. 
Just as in the case of the OBDM, the differences between the NQS and the exact diagonalization method increase with $A$ and are more visible for the more attractive values of $V_0$. In accordance with the observations of the local density profiles of Fig.~\ref{fig:density_profiles} and the OBDM of Fig.~\ref{fig:denmat}, the direct diagonalization results tend to have a more compact PCF, which also features some very faint
oscillations off the diagonal. We attribute these again to the truncation of the many-body basis in the exact diagonalization approach. 

\acknowledgments
This work is supported by STFC, through Grants No. ST/L005743/1 and No. ST/P005314/1; by Grants No. PID2020-118758GB-I00 and No. PID2020-114626GB-I00 funded by MCIN/AEI/10.13039/501100011033; 
by the ``Ram\'on y Cajal" Grant No. RYC2018-026072 funded by MCIN/AEI /10.13039/501100011033 and FSE “El FSE invierte en tu futuro”; 
by the ``Unit of Excellence Mar\'ia de Maeztu 2020-2023" award to the Institute of Cosmos Sciences, Grant No. CEX2019-000918-M funded by MCIN/AEI/10.13039/501100011033
and by the Generalitat de Catalunya, Grant No. 2021SGR01095. 
TRIUMF receives federal funding via a contribution agreement with the National Research Council of Canada.
This work has been financially supported by the Ministry of Economic Affairs and Digital Transformation of the Spanish Government through the QUANTUM ENIA project call – Quantum Spain project, and by the European Union through the Recovery, Transformation and
Resilience Plan – NextGenerationEU within the framework of the Digital Spain 2026 Agenda.
The authors acknowledge the computer resources at Artemisa, funded by the European Union ERDF and Comunitat Valenciana as well as the technical support provided by the Instituto de Física Corpuscular, IFIC (CSIC-UV).

\bibliographystyle{apsrev4-1}
\bibliography{biblio}

\end{document}